\pdfoutput=1

\documentclass[showpacs,aps,prl,twocolumn,superscriptaddress]{revtex4}
\usepackage{graphicx} 
\usepackage{dcolumn}
\usepackage{bm}
\usepackage{amssymb,amsmath}
\usepackage{epstopdf}
\usepackage[T1]{fontenc}
\usepackage[latin9]{inputenc}
\usepackage{color}
\usepackage{float}
\usepackage{graphicx}
\usepackage{esint}
\usepackage{bm}
\usepackage{graphics}
\makeatletter
\@ifundefined{textcolor}{}
{%
 \definecolor{BLACK}{gray}{0}
 \definecolor{WHITE}{gray}{1}
 \definecolor{RED}{rgb}{1,0,0}
 \definecolor{GREEN}{rgb}{0,1,0}
 \definecolor{BLUE}{rgb}{0,0,1}
 \definecolor{CYAN}{cmyk}{1,0,0,0}
 \definecolor{MAGENTA}{cmyk}{0,1,0,0}
 \definecolor{YELLOW}{cmyk}{0,0,1,0}
}

\begin{document}
\title{Dicke Superradiance in Solids}
\normalsize

\author{Kankan Cong}
\author{Qi Zhang}
\affiliation{Department of Electrical and Computer Engineering, Rice University, Houston, Texas 77005, USA}

\author{Yongrui Wang}
\affiliation{Department of Physics and Astronomy, Texas A\&M University, College Station, Texas 77843, USA}

\author{G. Timothy Noe II}
\affiliation{Department of Electrical and Computer Engineering, Rice University, Houston, Texas 77005, USA}

\author{Alexey Belyanin}
\affiliation{Department of Physics and Astronomy, Texas A\&M University, College Station, Texas 77843, USA}

\author{Junichiro Kono}
\thanks{Author to whom correspondence should be addressed}
\email[]{kono@rice.edu}
\affiliation{Department of Electrical and Computer Engineering, Rice University, Houston, Texas 77005, USA}
\affiliation{Department of Physics and Astronomy, Rice University, Houston, Texas 77005, USA}
\affiliation{Department of Materials Science and NanoEngineering, Rice University, Houston, Texas 77005, USA}

\date{\today}

\begin{abstract}
Recent advances in optical studies of condensed matter systems have led to the emergence of a variety of phenomena that have conventionally been studied in the realm of quantum optics. These studies have not only deepened our understanding of light-matter interactions but also introduced aspects of many-body correlations inherent in optical processes in condensed matter systems. This article is concerned with the phenomenon of superradiance (SR), a profound quantum optical process originally predicted by Dicke in 1954.  The basic concept of SR applies to a general $N$-body system where constituent oscillating dipoles couple together through interaction with a common light field and accelerate the radiative decay of the whole system.  Hence, the term SR ubiquitously appears in order to describe radiative coupling of an arbitrary number of oscillators in many situations in modern science of both classical and quantum description.  In the most fascinating manifestation of SR, known as superfluorescence (SF), an incoherently prepared system of $N$ inverted atoms spontaneously develops macroscopic coherence from vacuum fluctuations and produces a delayed pulse of coherent light whose peak intensity $\propto N^2$.  Such SF pulses have been observed in atomic and molecular gases, and their intriguing quantum nature has been unambiguously demonstrated.  In this review, we focus on the rapidly developing field of research on SR phenomena in solids, where not only photon-mediated coupling (as in atoms) but also strong Coulomb interactions and ultrafast scattering processes exist.  We describe SR and SF in molecular centers in solids, molecular aggregates and crystals, quantum dots, and quantum wells.  In particular, we will summarize a series of studies we have recently performed on semiconductor quantum wells in the presence of a strong magnetic field. In one type of experiment, electron-hole pairs were incoherently prepared, but a macroscopic polarization spontaneously emerged and cooperatively decayed, emitting an intense SF burst. In another type of experiment, we observed the SR decay of coherent cyclotron resonance of ultrahigh-mobility two-dimensional electron gases, leading to a decay rate that is proportional to the electron density.  These results show that cooperative effects in solid-state systems are not merely small corrections that require exotic conditions to be observed; rather, they can dominate the nonequilibrium dynamics and light emission processes of the entire system of interacting electrons.
\end{abstract}


\maketitle

\section{Introduction}
\label{intro}

\subsection{Dicke Phenomena}

The legacies of Robert H.\ Dicke (1916-1997) continue to influence many disciplines of modern physics, including cosmology, gravitation, atomic physics, condensed matter physics, and applied physics~\cite{HapperetAl97PT}. Although Dicke is likely to be best known for the development of the lock-in amplifier, he was also the inventor of a sensitive microwave receiver called the {\em Dicke radiometer}~\cite{Dicke46RSI}.  Dicke is also credited with proposing, in 1956, an open resonator design for amplifying infrared radiation~\cite{Dicke58Patent}, an essential component of lasers~\cite{Townes99Book}.  Dicke's theory of a collisional suppression of Doppler broadening ({\em Dicke narrowing})~\cite{Dicke53PR} is a crucial ingredient of atomic clocks currently mounted on GPS satellites.  Dicke and coworkers predicted~\cite{DickeetAl65AJ} the cosmic microwave background as a remnant of the Big Bang and started searching for it using a Dicke radiometer, only to become the second to Penzias and Wilson~\cite{PenziasWilson65AJ} (who also used a Dicke radiometer).  Dicke is also often cited as a central figure in the renaissance of gravitation and cosmology~\cite{Dicke57RMP,BransDicke61PR,Dicke61Nature,Dicke62Nature,Dicke70Book}, prolifically reporting innovative models, principles, and arguments that are now widely known under his name, including the {\em Brans-Dickey theory} of gravitation~\cite{BransDicke61PR}, the {\em Dicke anthropic principle}~\cite{Dicke61Nature,Dicke62Nature}, and the {\em Dicke coincidence}~\cite{Dicke70Book,Peebles93Book}.

Among these diverse ``Dicke phenomena'' found in various branches of physics, this article is concerned with a particular phenomenon called the {\em Dicke superradiance (SR)}~\cite{Dicke54PR}, by which Dicke introduced the profound concept of cooperative and coherent spontaneous emission. This general concept, as detailed below, has been studied in different areas of contemporary science and engineering, especially  quantum optics, condensed matter physics, optoelectronics, and plasmonics. Within the original context of atomic SR, many excellent review articles and monographs~\cite{Eberly72AJP,AndreevetAl80SPU,GrossHaroche82PR,ZheleznyakovetAl89SPU,ScullySvidzinsky09Science} and textbook chapters~\cite{SargentetAl74Book,AllenEberly75Book,Haken84Book,Siegman86Book,YamamotoImamoglu99Book} exist.

\subsection{Dicke Superradiance}
\label{SRSF-intro}

\begin{figure}[b]
\centering
\includegraphics[scale = 0.48]{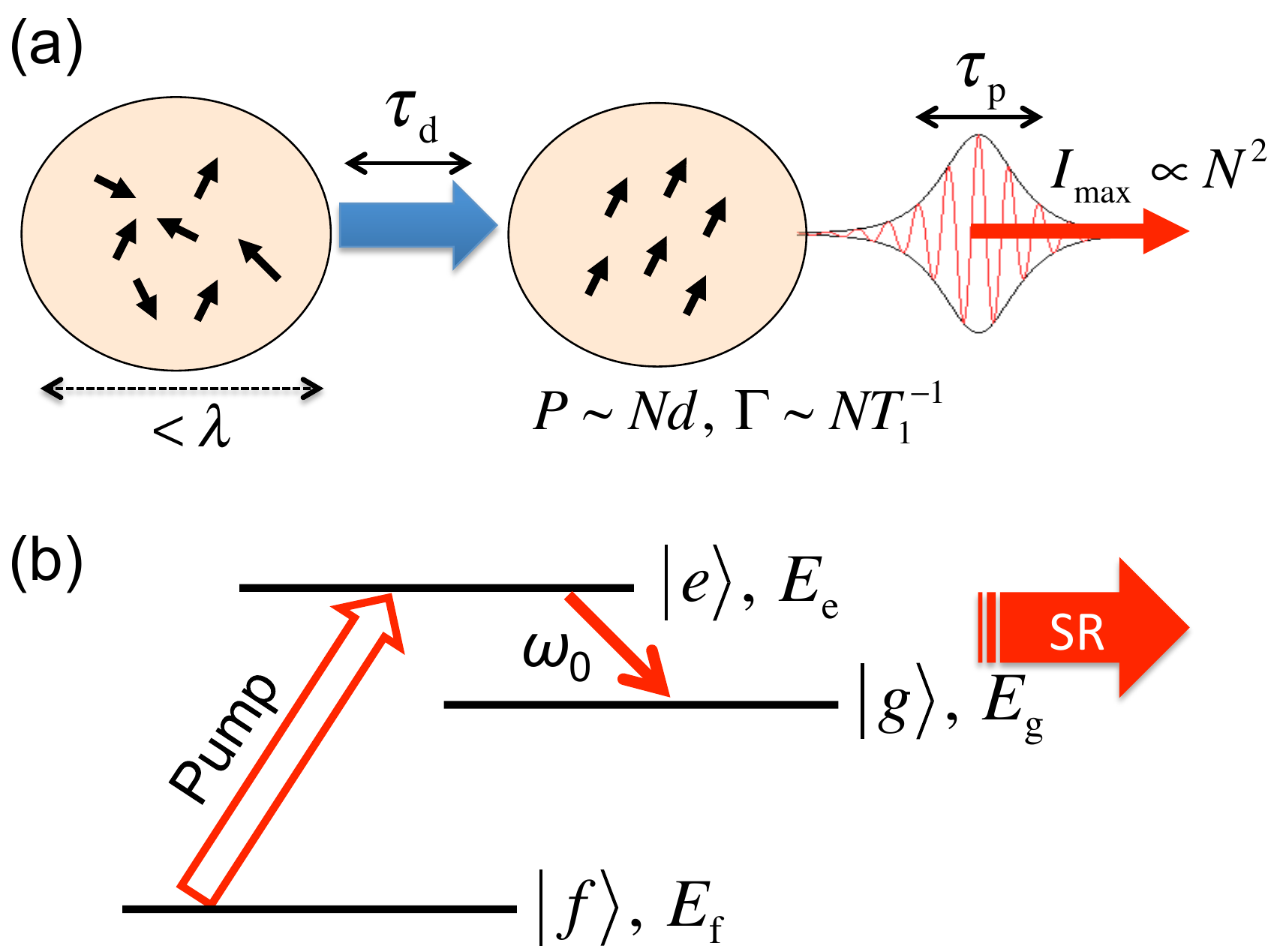}
\caption{(a)~Basic processes and characteristics of SF. An incoherent ensemble of $N$ excited two-level atoms is confined in volume $< \lambda^3$. At low densities, the spontaneous emission intensity $\propto N$ with decay rate $T_1^{-1}$. $T_1$: radiative decay time of an isolated atom.  At high densities, a giant dipole $P \sim Nd$ develops via photon exchange. $d$: individual atomic dipole moment. The $P$ decays at an accelerated rate $\Gamma_\textrm{SR} \sim N T_1^{-1}$ by emitting a pulse with peak intensity $I_{\rm max} \propto N^2$.  (b)~Typical level scheme for a SR experiment. Pulsed optical pumping of electrons from level $|f\rangle$ to level $|e\rangle$ creates population inversion between level $|e\rangle$ and level $|g\rangle$, leading to subsequent superradiance with photon frequency $\omega_{0}$ = $(E_\textrm{e} - E_\textrm{g})/\hbar$.}
\label{SRSchematic}
\end{figure}

In his pioneering paper in 1954~\cite{Dicke54PR}, Dicke studied the radiative decay of an ensemble of $N$ incoherently excited two-level atoms confined in a region of space with a volume $V$ smaller than $\sim$$\lambda^3$, where $\lambda$ is the wavelength corresponding to the photon energy equal to the level separation; see Fig.\,\ref{SRSchematic}(a). At low densities, the atoms do not interact with each other, and their spontaneous emission intensity $I_{\rm SE} \propto N$ with a decay rate given by $T_1^{-1}$, where $T_1$ is the spontaneous radiative decay time (population relaxation time) of an isolated atom.  At sufficiently high densities of inverted atoms, however, their dipole oscillations lock in phase through exchange of photons and develop a giant dipole $P \sim Nd$, where $d$ is the individual atomic dipole moment, over a characteristic delay time $\tau_\textrm{d}$. The macroscopic dipole decays at an accelerated rate $\Gamma_\textrm{SR} \sim N T_1^{-1}$ by emitting an intense coherent radiation pulse. The pulse duration $\tau_{\rm p} \propto 1/N$, so that the emitted light intensity scales as $I_{\rm SR} \propto N/\tau_{\rm p} \propto N^2$, a hallmark of coherent emission (``For want of a better term, a gas which is radiating strongly because of coherence will be called \lq{superradiant}\rq\,''~\cite{Dicke54PR}).

While Dicke did not distinguish between the terms SR and superfluorescence (SF), and in fact, never mentioned SF, subsequent studies (e.g.,~\cite{BonifacioLugiato75PRA,BonifacioLugiato75PRA2,VrehenetAl80Nature,VrehenGibbs82Book,SchuurmansetAl82AAMP}) have established the following semantic convention: i)~SR results when the coherent polarization is generated by an external coherent laser field, and ii)~SF occurs when the atomic system is initially incoherent and the macroscopic polarization develops {\em spontaneously} from quantum fluctuations; the resulting macroscopic dipole decays superradiantly at the last stage. In other words, SF emerges when there is no coherent polarization initially present in the system. The existence of this spontaneous self-organization stage makes SF a more exciting condensed matter subject but also much more difficult to observe than SR, especially in solids. In addition, SF is fundamentally a stochastic process: the optical polarization and the electromagnetic field grow from initially incoherent quantum noise to a macroscopic level. Thus, SF is \emph{intrinsically random}: even for identical preparation conditions, initial microscopic fluctuations get exponentially amplified and may result in macroscopic pulse-to-pulse fluctuations~\cite{HaakeetAl79PRL,VrehenetAl80Nature,RehlerEberly71PRA,BonifacioLugiato75PRA,BonifacioLugiato75PRA2,YouetAl91JOSAB}, e.g., in delay time $\tau_\textrm{d}$~\cite{YouetAl91JOSAB,FlorianetAl84PRA,AriunboldetAl12PLA}, pulse width~\cite{FlorianetAl84PRA}, and emission direction~\cite{JhoetAl06PRL}. For a detailed discussion on semantic confusion between SR and SF, see, e.g., pp.\,547-557 of Ref.~\cite{Siegman86Book}.

A typical scheme adopted in successful SF experiments is based on a three-level system, as shown in Fig.\,\ref{SRSchematic}(b). Initially, state $|f\rangle$ is fully occupied while states $|e\rangle$ and $|g\rangle$ are unoccupied.  At $t$ = 0, a short and intense pump laser pulse whose central frequency is resonant with the transition between $|f\rangle$ and $|e\rangle$ excites many atoms to state $|e\rangle$, thus producing a total population inversion between state $|e\rangle$ and state $|g\rangle$. A slightly more complicated configuration is realized when electrons are excited to higher states and then relax incoherently to state $|e\rangle$.  Note that in both scenarios there is no coherent macroscopic polarization present in the system on the $|e\rangle$ to $|g\rangle$ transition immediately after the pump pulse. The final stage, after a macroscopic polarization spontaneously develops, is the superradiant decay from state $|e\rangle$ to state $|g\rangle$, emitting an intense pulse with a central photon energy of $\hbar \omega_0$ = $E_\textrm{e} - E_\textrm{g}$.

Being coherent processes, SR and SF emerge only when the cooperative radiative decay of the system becomes faster than any other decoherence (phase breaking) processes. For example, SR is observable only when $\Gamma_\textrm{SR} \sim NT_1^{-1}$ is larger than any other scattering and relaxation rates. Requirements for observation of SF are more stringent.  SF pulses can develop only under the condition that both the pulse duration $\tau_\textrm{p}$ and the delay time $\tau_\textrm{d}$ can be made shorter than any phase breaking time scales, particularly, the population relaxation time, $T_1$, and the polarization relaxation time, $T_2$:
\begin{eqnarray}
\label{Eq::SR_time}
\tau_\textrm{p}, \tau_\textrm{d} < T_1, T_2.
\end{eqnarray}
Since $\tau_\textrm{p} \propto 1/N$ and $\tau_\textrm{d} \sim \tau_\textrm{p} \ln N$, achieving a large $N$, i.e., strong inversion, is crucial. Physically, a macroscopic (giant) polarization, $P$, must build up and decay in a time shorter than $T_2$ (which is usually much shorter than $T_1$ in solids).  Alternatively, the cooperative frequency $\Omega_\textrm{c}$~\cite{ZheleznyakovetAl89SPU,BelyaninetAl91SSC,BelyaninetAl92LP,BelyaninetAl97QSO,BelyaninetAl98QSO}, which determines the growth rate of the macroscopic polarization, has to be larger than the decoherence rate, i.e.,
\begin{align}
\label{coop}
\Omega_\textrm{c} = \sqrt{\frac{2\pi \tilde{\Gamma} \omega_{0}d^{2}\Delta n}{\tilde{n}_\textrm{op}^2 \hbar}} > \frac{1}{T_{1}}, \frac{1}{T_{2}}.
\end{align}
Here, $\Delta n$ is the population inversion density, $\tilde{\Gamma}$ is the overlap factor of the electromagnetic radiation mode with the active medium, and $\tilde{n}_\textrm{op}$ is the refractive index.  This is a necessary condition for SF, which is not easy to realize in solid-state systems using extended electronic states, as addressed in Section~\ref{Extended}.

Although Dicke's theory was purely quantum mechanical, some aspects of SR are classical in essence. Particularly, the aspect of synchronization and self-organization among oscillating dipoles, intrinsic in all SR processes, has many classical analogues, such as coupled pendulums~\cite{Huygens86Book}, metronomes~\cite{Pantaleone02AJP}, clapping hands~\cite{NedaetAl00Nature}, coupled plasmonic waveguides~\cite{Martin-CanoetAl10NL}, and an array of carbon nanotube antennas~\cite{RenetAl13PRB}. It is a natural consequence of electromagnetism that $N$ synchronized dipole oscillators, or antennas, radiate $N^2$ times more strongly. The essential idea of SR is that $N$ atoms behave as a {\em giant atom} and collectively decay with a rate that is $N$ times faster than that for an isolated atom. It must also be noted that essentially the same concept, known as {\em radiation damping (RD)}, was developed by Bloembergen and Pound~\cite{BloembergenPound54PR} in the context of magnetic resonance, also in 1954, independently of Dicke's work on SR~\cite{Dicke54PR}. Subsequent studies have firmly established the equivalence of RD and SR in a variety of systems~\cite{Eberly72AJP,Haken84Book,Bloom57JAP,Yariv60JAP,SandersetAl74PRB,ChiuetAl76SS,MatovetAl96JETP,Mikhailov04PRB,KrishnanMurali13PNMRS,ZhangetAl14PRL,LaurentetAl15PRL,AsadullinaetAl15arXiv}.

\begin{figure}[b]
\centering
\includegraphics[width=\linewidth]{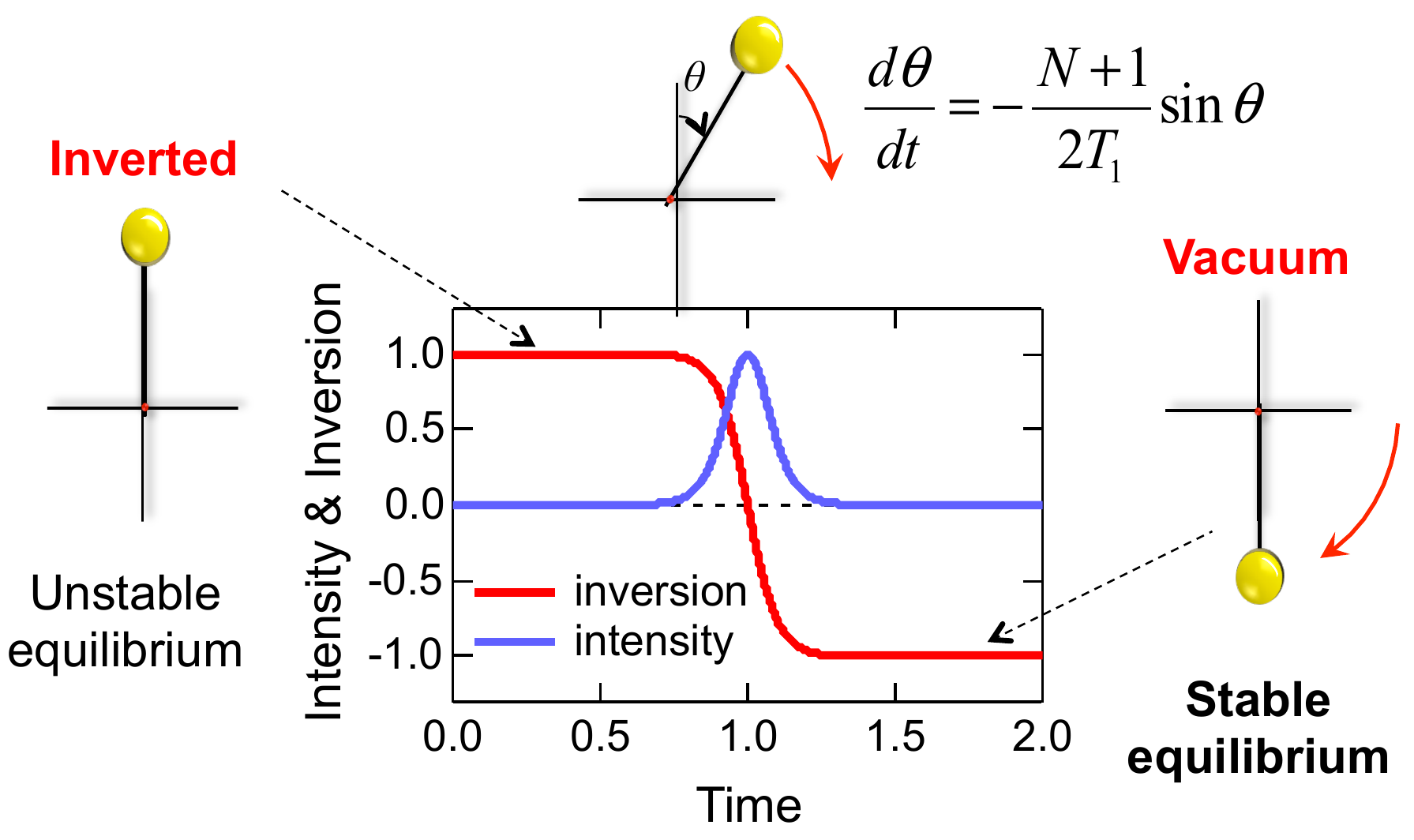}
\caption{Bloch vector representation of the SF emission process~\cite{NoeetAl12NP}.  The plots in the middle show the population inversion and emitted light intensity (normalized to the peak intensity) versus time (normalized to the pulse delay) for a SF system, with the dynamics of a Bloch vector dropping from the unstable excited state $\theta = 0^{\circ}$ to the ground state $\theta = 180^{\circ}$.}
\label{SFSchematic}
\end{figure}

Further analogies to pendulum motion can be used in visualizing the dynamics of a collective Bloch vector representing the $N$ two-level atoms during SF, as shown in Fig.\,\ref{SFSchematic}. At the initial stage, where all the dipoles are prepared in the excited state and there is no definite phase relationship among them, the Bloch vector points to `north'; this is an unstable equilibrium position for an inverted pendulum. Once the emission process starts, the Bloch vector will tend to drop towards the ground state where it points `south.' The equation of motion for the Bloch vector is indeed equivalent to that for a classical pendulum~\cite{Dicke54PR,AndreevetAl80SPU,RehlerEberly71PRA,Eberly72AJP,NoeetAl12NP}
\begin{align}
\label{Eq::SF}
\frac{d\theta}{dt}=\frac{-(N+1)\sin\theta}{2T_{1}},
\end{align}
where $\theta$ is the angle of the Bloch vector direction with respect to the vertical axis.  Equation~(\ref{Eq::SF}) indicates that the rate of change of $\theta$ is proportional to $\sin\theta$, and thus, once it starts moving, the motion gets faster with increasing $\theta$ at the beginning, and reaches the fastest rate at $\theta$ = 90$^{\circ}$, and then gradually slows down; it finally stops at $\theta$ = 180$^{\circ}$, where the Bloch vector points `south,' all the dipoles are in the ground state, and the population inversion is zero, i.e., all the energy in this system has been transferred to light through SF emission. Therefore, SF converts all energy stored in an inverted system into radiation, in contrast to {\em amplified spontaneous emission}, in which no more than half of the initial energy is consumed by the radiation pulse.

The above pendulum analogy (Fig.\,\ref{SFSchematic}) also illuminates the intrinsically quantum nature of SF.  Namely, when the Bloch vector is stable at the north pole ($\theta$ = 0$^\circ$), classically, it must stay there forever in the absence of any external perturbation.  However, fluctuations (quantum noise) of vacuum induce a finite tilt, which makes $d\theta/dt$ finite, initiating the whole process of macroscopic polarization buildup and collective radiative decay.  The amount of such quantum-fluctuation-driven initial tipping has been calculated~\cite{RehlerEberly71PRA,BonifacioLugiato75PRA,BonifacioLugiato75PRA2,MacGillivrayFeld76PRA} and measured~\cite{VrehenSchuurmans79PRL} in atomic ensembles. The macroscopic polarization thus starts from noise and builds up through photon exchange.  

The distinctive feature of SF is that the system of initially uncorrelated $N$ dipole oscillators evolves into a correlated superradiant state, where individual dipoles are oscillating in phase and contribute constructively to radiation. In principle, there also exists a {\em subradiant} state, which is also highly correlated but in which individual dipoles are out of phase and interfere destructively~\cite{Dicke54PR}. As a result, the net polarization in this state is greatly reduced. Such states cannot be formed through the development of SF starting from initially inverted identical quantum dipoles. However, if the dipoles are not identical, such as those in a system of inhomogeneously broadened two-level atoms, subradiant and superradiant states can coexist~\cite{TemnovWoggon05PRL}. Subradiant and superradiant correlations have recently been predicted to affect the lasing threshold for coupled quantum-dot nanolasers~\cite{LeymannetAl15PRA}. Also, subradiant states can be accessed in a system of a few degrees of freedom, where they are separated in energy from superradiant states. For example, in a compound plasmonic disc/ring nanocavity, two partial plasmonic modes are hybridized into a superradiant state with two dipole oscillations locked in phase and a subradiant state with two dipole oscillations out of phase by $\pi$~\cite{SonnefraudetAl10ACSNano}. They are observed as broad and narrow absorption resonances, well separated in energy.

Finally, it should be noted that a simple scaling law of the SR/SF intensity $\propto N^2$ is valid only for a Dicke model of a small atomic sample ($V \ll \lambda^3$). This condition can be safely met only in microwave experiments, but not in experiments performed in the infrared and visible ranges.  In most of the successful SR/SF experiments carried out to date, one or more dimensions of the active sample under study were much greater than the wavelength of emitted light.  Therefore, propagation, diffraction, defects, and fluctuations play major roles, and the emitted SF pulse undergoes strong and complicated nonlinear interactions with the inverted medium while propagating~\cite{ErnstStehle68PR,Ponte-GoncalvesetAl69PR,Lehmberg70PRA,Lehmberg70PRA2,Agarwal70PRA,ArecchiCourtens70PRA,BonifacioetAl71PRA,BonifacioetAl71PRA2,RehlerEberly71PRA,BonifacioLugiato75PRA,BonifacioLugiato75PRA2,MacGillivrayFeld76PRA,GrossHaroche82PR,PrasadGlauber00PRA}.  The shape of the pulse can significantly change through propagation and diffraction, and ringing can occur due to the coherent nature of interaction with the medium~\cite{HeinzenetAl85PRL}, all of which have to be taken into account through the Maxwell and Bloch equations to correctly explain experimental details.



\subsection{Dicke Phase Transition}


A general Hamiltonian of a system of $N$ two-level atoms dipole-coupled to a quantized radiation field is written as
\begin{align}
\label{Eq::SR_H}
 \hat{H} = \sum_{j=1}^{N}\left [\frac{1}{2m}\left \{\vec{p_{j}}-\frac{e}{c}\vec{A}(\vec{r_{j}})\right \}^{2}+U(\vec{r_{j}})\right ]+\hbar\omega a^{\dag}a,
\end{align}
where $\vec{A}$ is the vector potential representing the radiation field and $a$ and $a^{\dag}$ denote, respectively, the photon annihilation and creation operators.  In Dicke's model~\cite{Dicke54PR}, the following assumptions were made: (a)~The long-wavelength limit ($V \ll \lambda^3$) allows the vector potential to be evaluated at the center common to all atoms $\vec{A}(\vec{r_{j}})\simeq \vec{A}(0)$; (b)~the $A^{2}$ term is negligibly small; 
and (c)~the rotating-wave approximation (RWA) is valid. 
With these assumptions made, Eq.~(\ref{Eq::SR_H}) can be simplified to
\begin{align}
\label{Eq::SR_H1}
 \hat{H}_\textrm{Dicke} = \frac{\hbar\omega_{ba}}{2}\sum_{j=1}^{N}\sigma_{j}^{z}+\hbar\omega a^{\dag}a+\frac{\Lambda}{\sqrt{N}}\sum_{j=1}^{N}(\sigma_{j}^{+}a+\sigma_{j}^{-}a^{\dag}),
\end{align}
where the coupling constant $\Lambda \equiv \omega_{ba}d_{ba}(2\pi\hbar/\omega)^{1/2}(\rho)^{1/2}$, $d_{ba}$ is the dipole moment of the transition, $\rho$ $=$ $N/V$ is the atomic density, and $\sigma_{j}^{z}$, $\sigma_{j}^{+}$, and $\sigma_{j}^{-}$ are Pauli matrices used to describe the $j$-th atom. $\hat{H}_\textrm{Dicke}$ is referred to as the {\em Dicke Hamiltonian}.

In 1973, based on the Dicke Hamiltonian, Hepp and Lieb~\cite{HeppLieb73AP} calculated the free energy of the system exactly in the thermodynamic limit, showing that the system exhibits a second-order phase transition from a normal state to a {\em superradiant phase} at a certain critical temperature, $T_{\rm c}$, when the light-matter coupling strength, $\Lambda$, is sufficiently large. Also in 1973, Wang and Hioe~\cite{WangHioe73PRA} independently came to the same conclusion by calculating the canonical partition function.  The transition, which has come to be known as the {\em Dicke phase transition (DPT)}, occurs under strong coupling, $2\Lambda > \hbar \omega_{ba}$, at $T_{\rm c}$ derived from
\begin{align}
\label{Eq::SR_Tc}
\frac{(\hbar \omega_{ba})^2}{4 \Lambda^{2}} = \tanh\left (\frac{1}{2}\frac{\hbar \omega_{ba}}{k_{\rm B}T_\text{c}}\right ),
\end{align}
where $k_{\rm B}$ is the Boltzmann constant and the resonance $\omega_{ba} = \omega$ has been assumed for simplicity. It was also confirmed that this phase transition persists even without the RWA~\cite{CarmichaeletAl73PLA}.

\begin{figure}[b]
\centering
\includegraphics[scale=0.53]{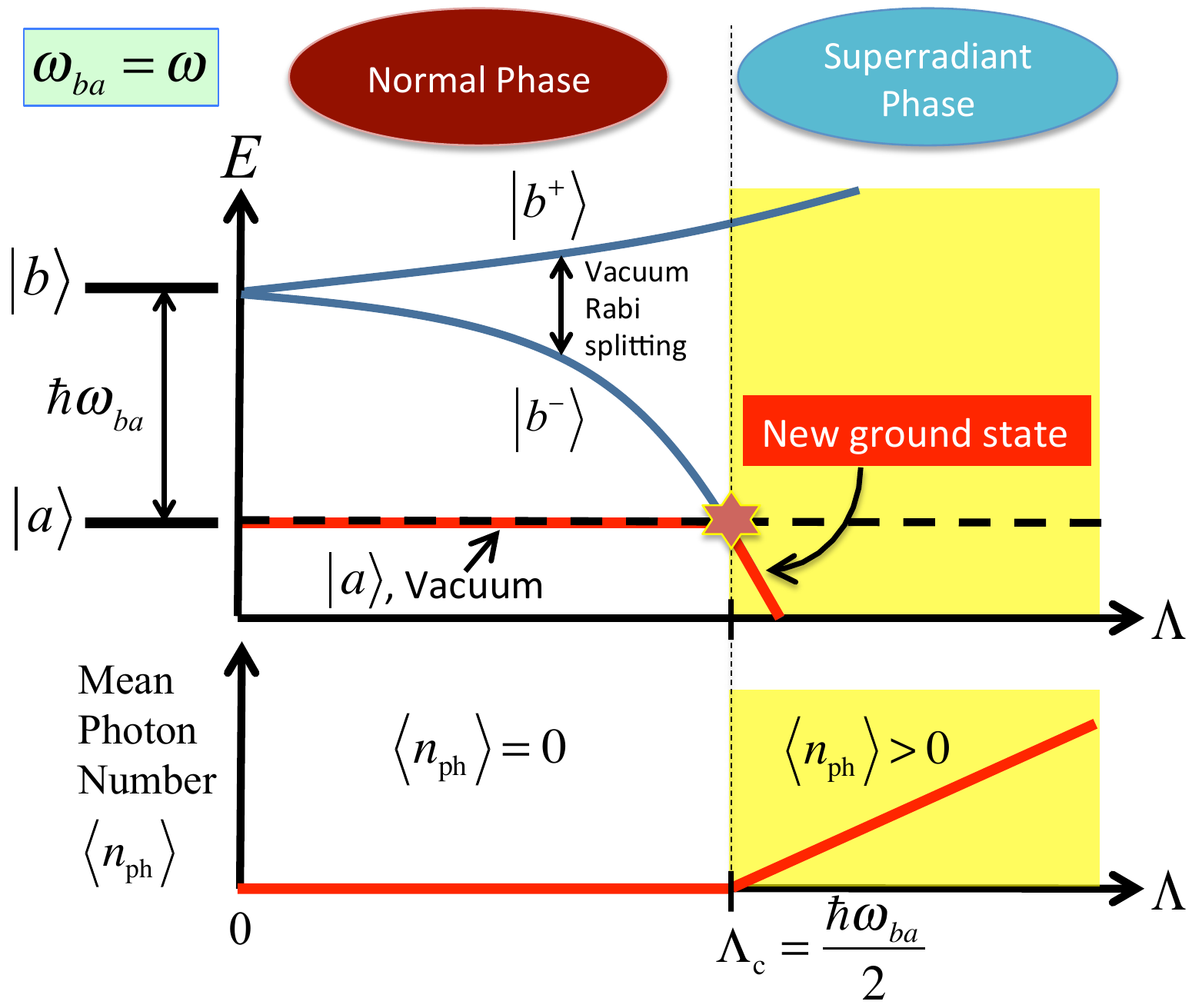}
\caption{Appearance of the superradiant phase, when the light-matter coupling constant exceeds the critical value, $\Lambda_\text{c}$.}
\label{DPT}
\end{figure}

However, such a prediction was soon challenged by Rza\.{z}ewski, W\'odkiewicz, and \.{Z}akowicz~\cite{RzazewskietAl75PRL}, who demonstrated that the presence of the DPT is entirely due to the neglect of the $A^{2}$ term [assumption (b) above]. When this term is included, the Dicke Hamiltonian becomes
\begin{equation}
\label{Eq::SR_H2}
\begin{split}
\hat{H}_\textrm{RW\.Z} = &\frac{\hbar\omega_{ba}}{2}\sum_{j=1}^{N}\sum_{j}^{z}+\hbar\omega a^{\dag}a +\\
      & \frac{\Lambda}{\sqrt{N}}\sum_{j=1}^{N}(\sigma_{j}^{+} + \sigma_{j}^{-})(a+a^{\dag}) + \kappa(a+a^{\dag})(a+a^{\dag}),
\end{split}
\end{equation}
where $\kappa \equiv \frac{e^{2}}{2m}\frac{2\pi\hbar}{\omega}\rho$. If $\hat{H}_\textrm{RW\.Z}$ is used, the finite-$T$ classical phase transition disappears in the case of electric dipole coupling. More recently, it has been shown that the Dicke Hamiltonian exhibits a quantum phase transition (QPT)~\cite{EmaryBrandes03PRL,LambertAl04PRL,BuzeketAl04PRL}, which occurs at $T =$ 0 as a function of $\Lambda$; above the critical coupling constant, $\Lambda_\text{c} = \hbar\omega_{ba}/2$, a superradiant phase appears, where the mean photon number, $\langle n_\text{ph}\rangle$, is finite (Fig.\,\ref{DPT}).
Nataf and Ciuti~\cite{NatafCiuti10NC} showed that, when the $A^{2}$ term is included, the QPT vahishes for cavity quantum electrodynamics (QED) systems but still persists for circuit QED systems, where the wave function of a Cooper pair is different from an atomic wave function, which is limited by the oscillator strength sum rule.

The DPT has since been discussed in a variety of situations~\cite{EasthamLittlewood01PRB,CiutietAl05PRB,BaumannetAl10Nature,HagenmullerCiuti12PRL,ChirollietAl12PRL,ChenetAl14PRL}. Ciuti {\it et al}.\ considered a system in which a microcavity photon mode is strongly coupled to a semiconductor intersubband transition, showing that tuning quantum properties of the ground state by changing the Rabi frequency via an electrostatic gate can bring the system into the strong coupling regime, where correlated photon pairs can be generated~\cite{CiutietAl05PRB}. Experimentally, a DPT has been realized in an open system formed by a Bose-Einstein condensate coupled to an optical cavity by observing the emergence of a self-organized supersolid phase, which is driven by infinitely long-range interactions between the condensed atoms induced by a two-photon process involving the cavity mode and a pump field~\cite{BaumannetAl10Nature}. By increasing the pump power over time while monitoring the light leaking out of the cavity, the self-organization behavior can be observed, in the sense that a critical pump power leads to an abrupt increase in the mean intracavity photon number. In contrast with a Bose gas, superradiance from a degenerate Fermi gas in a cavity is theoretically predicted to be enhanced due to the Fermi surface nesting effect, thus it can be achieved with a much smaller critical pumping field strength~\cite{ChenetAl14PRL}. 

Recently, the possibility of realizing a Dicke phase transition in a graphene cavity QED system has been discussed theoretically by Hagenm\"{u}ller {\it et al}.~\cite{HagenmullerCiuti12PRL} and Chirolli {\it et al}.~\cite{ChirollietAl12PRL}, who reached opposite conclusions. Hagenm\"{u}ller {\it et al}.\ argued that, by putting graphene in a perpendicular magnetic field, the ultrastrong coupling regime characterized by a vacuum Rabi frequency comparable or even larger than the transition frequency can be obtained for high enough filling factors of the graphene Landau levels. Due to the linear conical dispersion at low energies, the role of the $A^{2}$ term can be negligible when the lattice constant is much smaller than the magnetic length, thus allowing the possibility of a Dicke phase transition in a graphene cavity QED system for a large electron density. Chirolli {\it et al}., on the other hand, emphasized the importance of the $A^{2}$ term in the strong coupling regime, which is dynamically generated by interband transitions, and concluded that the Dicke phase transition is forbidden in such a system~\cite{ChirollietAl12PRL}.


\section{Superfluorescence Observations in Atomic and Molecular Gases}
\label{Gases}

\begin{figure}[h]
\centering
\includegraphics[scale=0.47]{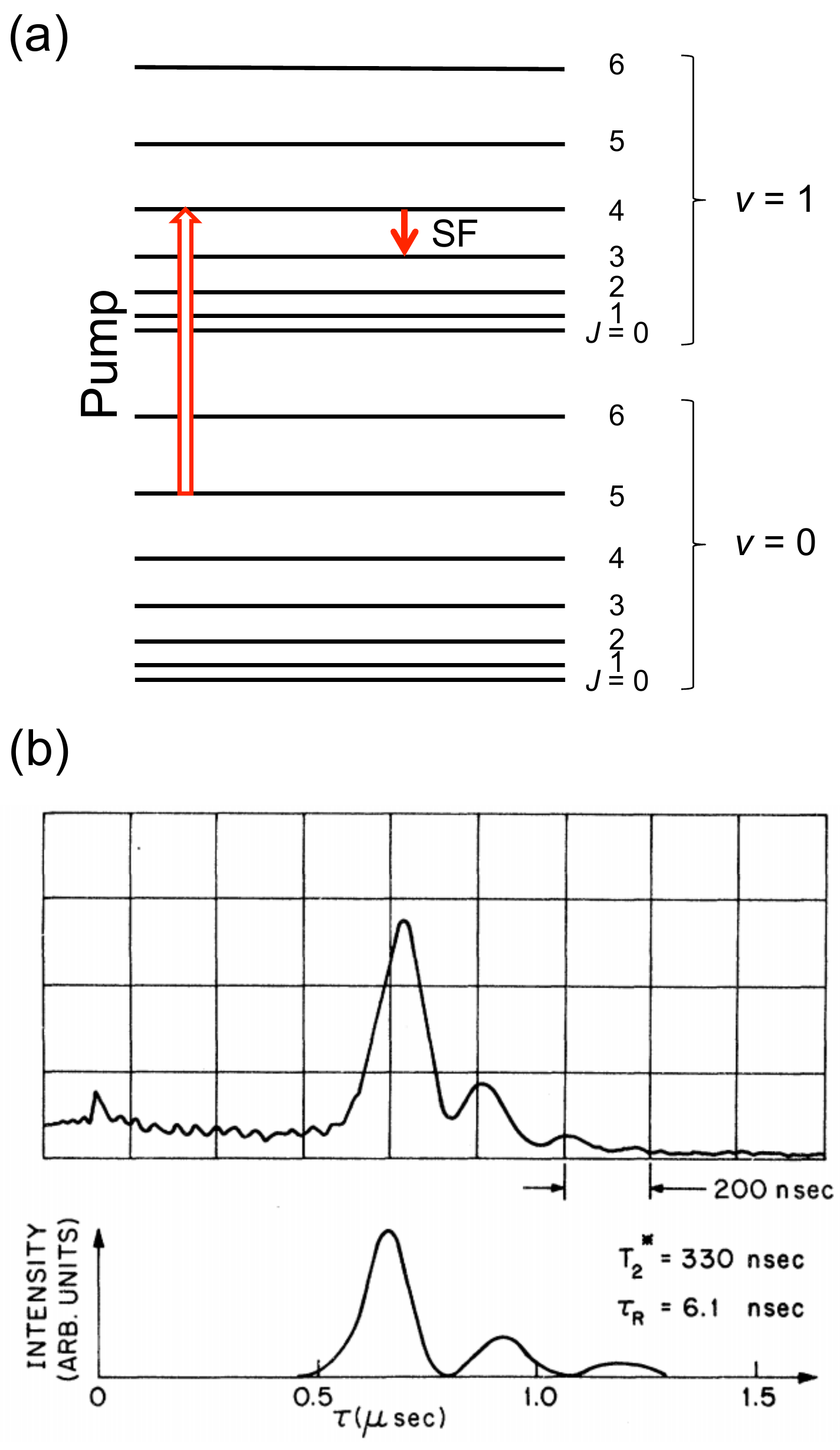}
\caption{First observation of superfluorescence (SF) by Skribanowitz {\it et al}.\ using a gas of HF molecules. (a)~Energy-level diagram of a HF molecule with the pump and the superradiant (SR) transitions indicated by arrows [see also Fig. 1(b)]. Reproduced (adapted) with permission from~\cite{AndreevetAl80SPU}. Copyright 1980, American Institute of Physics. (b) Oscilloscope trace (top trace) of an observed SF pulse at 84\,$\mu$m from a HF gase at a pressure of 1.3\,mTorr, pumped by an HF laser beam at the $R_{2}(2)$ line with a peak intensity of 1\,kW/cm$^2$, together with a theoretical fit (bottom trace) using coupled Maxwell-Schr\"{o}dinger equations. Reproduced (adapted) with permission from~\cite{SkribanowitzetAl73PRL}. Copyright 1973, American Physical Society.}
\label{HFSF}
\end{figure}

The first experimental observation of SF was made in a gas of hydrogen fluoride (HF) molecules by Skribanowitz and coworkers in 1973~\cite{SkribanowitzetAl73PRL}. 
They pumped the gas with a laser beam at 3\,$\mu$m to excite the molecules from one of the rotational sublevels in the ground vibrational state ($v$ = 0) to one of the rotational sublevels in the first vibrational state ($v$ = 1); see Fig.\,\ref{HFSF}(a). Under strong enough pumping, this excitation scheme produced a complete population inversion between two sublevels, the ($J$ + 1)-th and $J$-th rotational levels, within the $v$ = 1 state, which generated a delayed pulse with a far-infrared wavelength. 
The top trace in Fig.\,\ref{HFSF}(b) is an observed SF pulse at 84\,$\mu$m, corresponding to the $J$ = 3 $\rightarrow$ 2 transition, with a pulse width of $\sim$20\,ns and a delay time of $\sim$700\,ns. In addition to the first, main pulse, trailing ringing was also observed. The intensity, pulse width, and delay time of the main SF pulse changed with the pump intensity and gas pressure in manners qualitatively consistent with theoretical expectations~\cite{ArecchiCourtens70PRA,BonifacioetAl71PRA,BonifacioetAl71PRA2}.  Particularly, as the pressure or pump intensity was reduced, the pulse delay and width increased while the emission amplitude decreased.  The authors were able to fit their experimental data using a semiclassical model based on coupled Maxwell-Schr\"{o}dinger equations, which was later elaborated~\cite{MacGillivrayFeld76PRA}.  Adjusting parameters, they were able to reproduce both the main pulse and ringing; see the bottom trace in Fig.\,\ref{HFSF}(b).  Later, Heizen and coworkers~\cite{HeinzenetAl85PRL} demonstrated that ringing is an intrinsic property of SF, reflecting the coherent Rabi-type interaction of the propagating SF pulse with the medium (termed the {\em Burnham-Chiao ringing}~\cite{BurnhamChiao69PR}).

Subsequent experiments observed SF in different gas species and in different wavelength ranges~\cite{GrossetAl76PRL,FlusbergetAl76PLA,KumarakrishnanetAl05JOSAB}.  Gross {\it et al}.\ observed SF in the mid-infrared rage (2.21, 3.41, and 9.10\,$\mu$m) from a gas of atomic sodium~\cite{GrossetAl76PRL}.  In this short-wavelength range, the dephasing process due to the Doppler effect was much faster than that in the above far-infrared experiment by Skribanowitz and coworkers, and consequently, the observed SF pulse widths were much narrower (in the nanosecond range).
In addition, Flusberg and coworkers used the 7 $^2$P$_{1/2}$ $\rightarrow$ 7 $^2$S$_{1/2}$ transition in a vapor of atomic thallium and observed SF at 1.30\,$\mu$m; superradiant delays of up to 12\,ns were observed~\cite{FlusbergetAl76PLA}.  Similarly to the experiment by Skribanowitz {\it et al}.\ in HF~\cite{SkribanowitzetAl73PRL}, these experiments also observed coherent ringing.

\begin{figure}[h]
\centering
\includegraphics[scale=1]{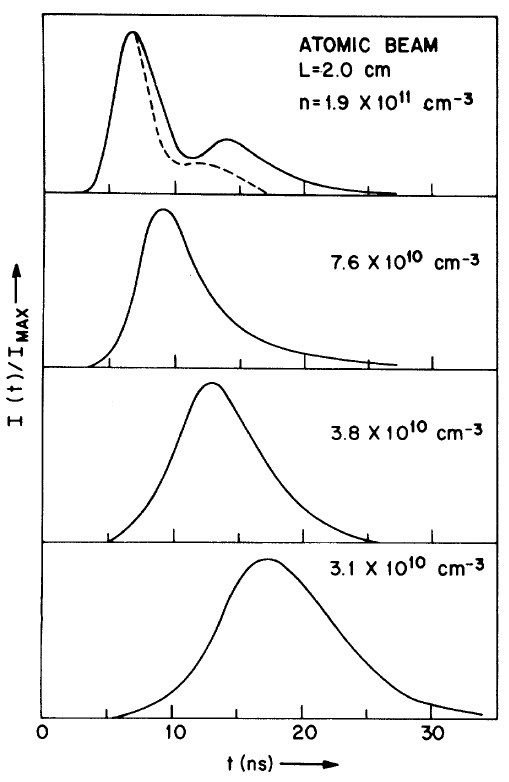}
\caption{Single-shot SF experiments in Cs vapor by Gibbs {\it et al}.~\cite{GibbsetAl77PRL}. Single-pulse (ringing-free) SF is observed under the conditions for ``pure'' SF~\cite{BonifacioLugiato75PRA,BonifacioLugiato75PRA2}. Reproduced with permission from~\cite{GibbsetAl77PRL}. Copyright 1977, American Physical Society.}
\label{GibbsSF}
\end{figure}

Gibbs {\it et al}.~\cite{GibbsetAl77PRL} observed ringing-free, single-pulse SF in cesium (Cs) gas under the conditions specified by Bonifacio and coworkers~\cite{BonifacioLugiato75PRA,BonifacioLugiato75PRA2} for "pure" SF to be observable: (a)~a pure two-level system; 
(b)~the Fresnel number $F$ = $A/\lambda L$ $\approx$ 1, where $A$ and $L$ are the cross-sectional area and length of a pencil-shaped sample, respectively; 
and (c)~$\tau_\textrm{e} < \tau_\textrm{c} < \tau_\textrm{p} < \tau_\textrm{d} < T_{1}, T_{2}, T_{2}^{*}$ and $\tau_\textrm{P} \ll \tau_\textrm{d}$, where $\tau_\textrm{e} = L/c$, $\tau_\textrm{c} = (\tau_\textrm{e} \tau_\textrm{p})^{1/2}$, $\tau_\textrm{p} = 8\pi T_1/3\rho\lambda^2 L$, $\rho$ is the number density of atoms, 
and $\tau_\textrm{P}$ is the pump pulse width. 
Under these conditions, it was found that single pulses can be observed for delay times beyond 7\,ns. For shorter delay times, multiple pulses occur with shapes fluctuating greatly from pulse to pulse even at the same delay time. 


\section{Cooperative Spontaneous Emission from Atomic-Like States in Solids}
\label{Localized}


This section concerns observations of superradiant emission in solids utilizing {\em localized} states: molecular centers in solids, molecular aggregates/crystals, semiconductor quantum dots and nanocrystals.  These systems are atomic-like, i.e., they retain most of the properties of a dilute ensemble of two-level atoms. The radiative transition in these systems is typically shielded from decoherence processes in the solid matrix, and the density of active dipoles is sufficiently low so that any nonradiative interaction between them is negligible.  Therefore, all the concepts and methodologies developed earlier for atomic SR and SF directly apply to these systems. However, the solid-state environment enables experiments that are difficult for gaseous samples (e.g., temperature dependence) and opens up new device application possibilities for developing light sources based on cooperative spontaneous emission. 

\subsection{Molecular Centers in Solids}
\label{MolecularCenters}

\begin{figure}[hbpt]
\centering
\includegraphics[width=0.75\linewidth]{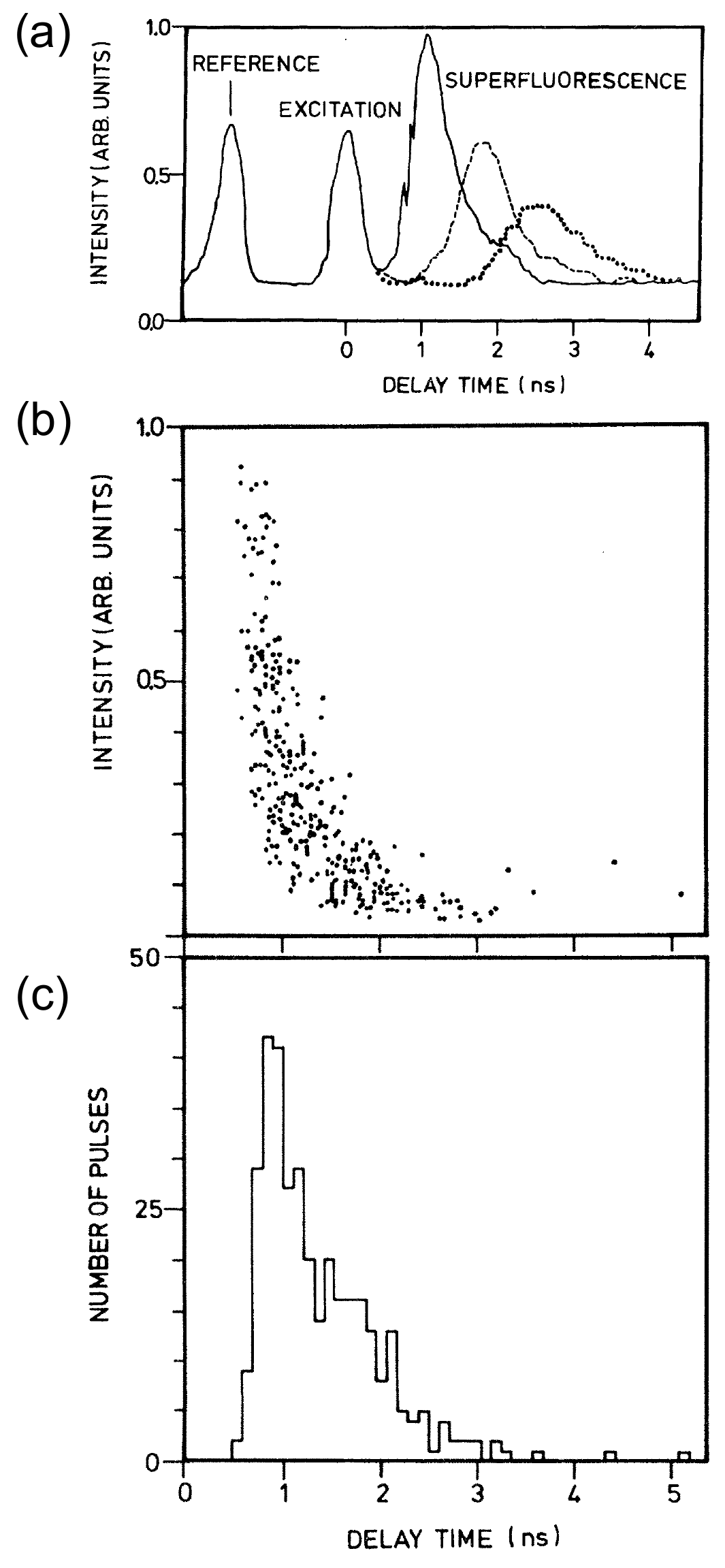}
\caption{Superfluorescence observed in a KCl crystal containing O$_{2}^{-}$ centers~\cite{FlorianetAl82SSC,FlorianetAl84PRA}.  (a)~SF pulses observed under different conditions. (b)~Intensity of observed SF pulses versus delay time, showing the tendency that the intensity is higher when the delay time is shorter.  (c)~Histogram of observed delay times for 300 SF pulses. Reproduced (adapted) with permission from~\cite{FlorianetAl84PRA}. Copyright 1984, American Physical Society.}
\label{KClO2SF}
\end{figure}

The first observation of SF in solids was made by Florian, Schwan, and Schmid in crystals of oxygen-doped alkali halide, KCl:O$_{2}^{-}$, at low temperatures ($<$30\,K).  A preliminary report on the observation in 1982~\cite{FlorianetAl82SSC} was substantiated by a detailed subsequent study in 1984~\cite{FlorianetAl84PRA}. They used ultraviolet pulses (265~nm, pulse duration $\sim$ 30\,ps, peak intensity $\sim$ 10\,GW/cm$^{2}$) from a frequency-quadrupled mode-locked Nd-YAG laser to excite the crystal and observed SF pulses with visible wavelengths (592.8\,nm and 629.1\,nm) in the time domain; see Fig.\,\ref{KClO2SF}(a).  
Around the same time, Zinov'ev {\it et al}.\ reported possible SR in a diphenyl crystal containing pyrene molecules at 4.2\,K~\cite{ZinovievetAl83JETP}. 
They excited pyrene molecules with the third harmonic of a Y$_{3}$Al$_{5}$O$_{12}$:Nd laser and observed emission at 373.9\,nm. 
Above a threshold excitation intensity, a drastic reduction in the radiative decay time occurred (from 110\,ns to 5-6\,ns). 
At the same time, the emission was highly directional (solid angle $\sim$ 0.1\,sr) and the linewidth increased with the pump intensity, suggesting that the observed emission in this regime was due to SR or SF.  However, delayed pulses, expected for SF, were not observed.

In the time-domain experiments by Florian, Schwan, and Schmid~\cite{FlorianetAl84PRA}, the intrinsically random nature of SF, as discussed in Section~\ref{SRSF-intro}, was clearly demonstrated. Even under identical excitation conditions, the intensity, pulse width, and delay time of SF were found to vary strongly from shot to shot.  The pulse intensities fluctuated by more than a factor of 10, while the pulse width varied between 0.5 and 6\,ns and the pulse delay time changed between 0.5 and 10\,ns. Figure~\ref{KClO2SF} summarizes their analysis of 300 individual SF pulses for which the excitation conditions were identical.  Figure~\ref{KClO2SF}(b) plots the pulse intensity against delay time, which shows the correlation between the two that for a shorter time delay the intensity is higher, while Fig.\,\ref{KClO2SF}(c) is a histogram of the observed delay times.


\begin{figure}[hbtp]
\centering
\includegraphics[scale=1]{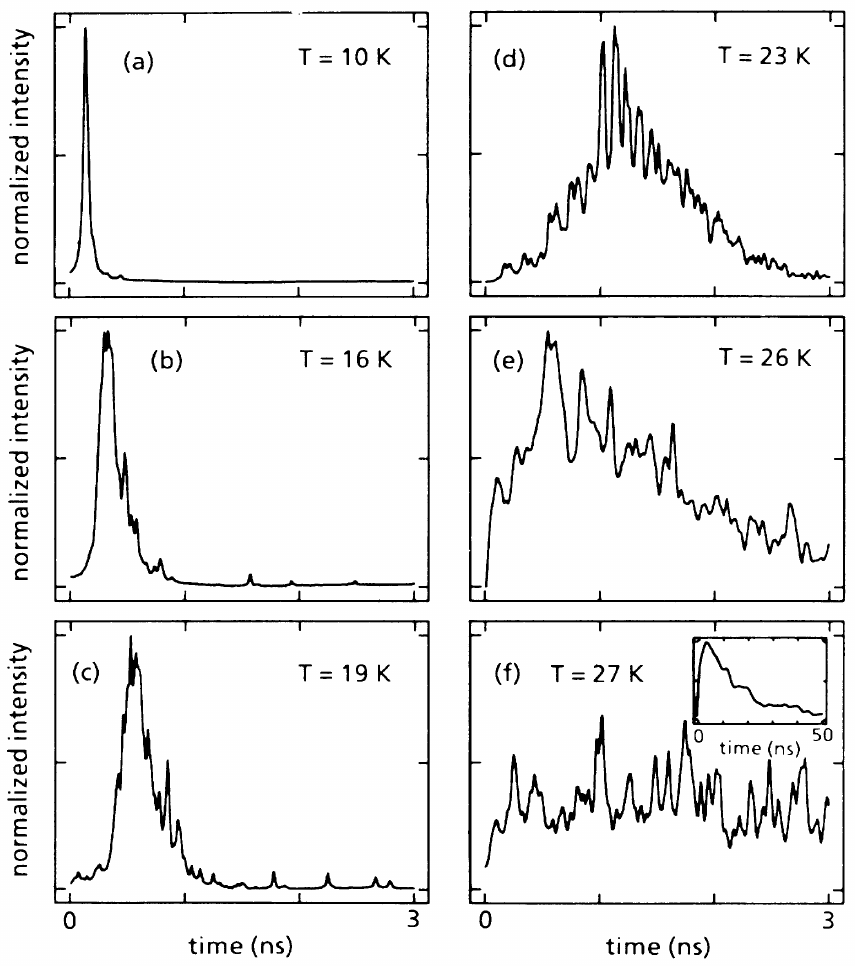}
\caption{Observed transition from superfluorescence to amplified spontaneous emission in an O$_{2}^{-}$-doped KCl crystal as a function of temperature from 10\,K to 27\,K~\cite{MacuitetAl87PRL}. Reproduced with permission from~\cite{MacuitetAl87PRL}. Copyright 1987, American Physical Society.}
\label{SF-to-ASE}
\end{figure}

In a subsequent study on KCl:O$_{2}^{-}$ by Macuit {\it et al}., a transition from SF to amplified spontaneous emission (ASE) was observed~\cite{MacuitetAl87PRL}.  The transition occurred as the dephasing time increased from 10$^{-10}$\,s to 3 $\times$ 10$^{-11}$\,s as the lattice temperature increased from 10\,K to 27\,K.  As shown in Fig.\,\ref{SF-to-ASE}, at 10\,K a sharp SF pulse was observed at a time delay of 160\,ps with a pulse width of 60\,ps. As the temperature was increased, the pulse broadened and the peak intensity dropped.  The time delay initially increased but then began to decrease, and at the highest temperature (27\,K), there was essentially no time delay.  These observations are consistent with the prediction~\cite{SchuurmansPolder79PLA} that the emission is characteristic of SF if $T_2 > (\tau_\textrm{p} \tau_\textrm{d})^{1/2}$, and is characteristic of ASE if $(\tau_\textrm{p} \tau_\textrm{d})^{1/2} > T_2 > \tau_\textrm{p}$.  Here, $\tau_\textrm{p} = 8\pi T_1/3 \rho \lambda^3 L$ is the pulse width, $\tau_\textrm{d} = \tau_\textrm{p} [ \ln (2\pi N)^{1/2} ]^2/4$ is the time delay, $\rho$ is the number density of atoms, $L$ is the length of the sample, and $N$ is the total number of atoms.


\subsection{Molecular Aggregates and Crystals}

Cooperative spontaneous emission processes have also been investigated in {\em molecular solids}, such as molecular aggregates and crystals, including {\it J}-aggregates~\cite{DeBoerWiersma90CPL,FidderetAl90CPL}, LH-2 photosynthetic antenna complexes~\cite{MonshouweretAl97JPCB}, $\pi$-conjugated polymer thin films~\cite{FrolovetAl97PRL,KhachatryanetAl12PRB}, {\it H}-aggregates~\cite{Spano00CPL,MeinardietAl03PRL}, and tetracene thin films and nanoaggregates~\cite{LimetAl04PRL}.  In all these studies, accelerated radiative decay was observed and attributed to cooperative emission, although SF, in the form observed in atomic systems (Section~\ref{Gases}) and molecular centers in crystals (Section~\ref{MolecularCenters}), has not been reported and some of the reported results and claims remain controversial.

\begin{figure}[hbtp]
\centering
\includegraphics[scale=0.67]{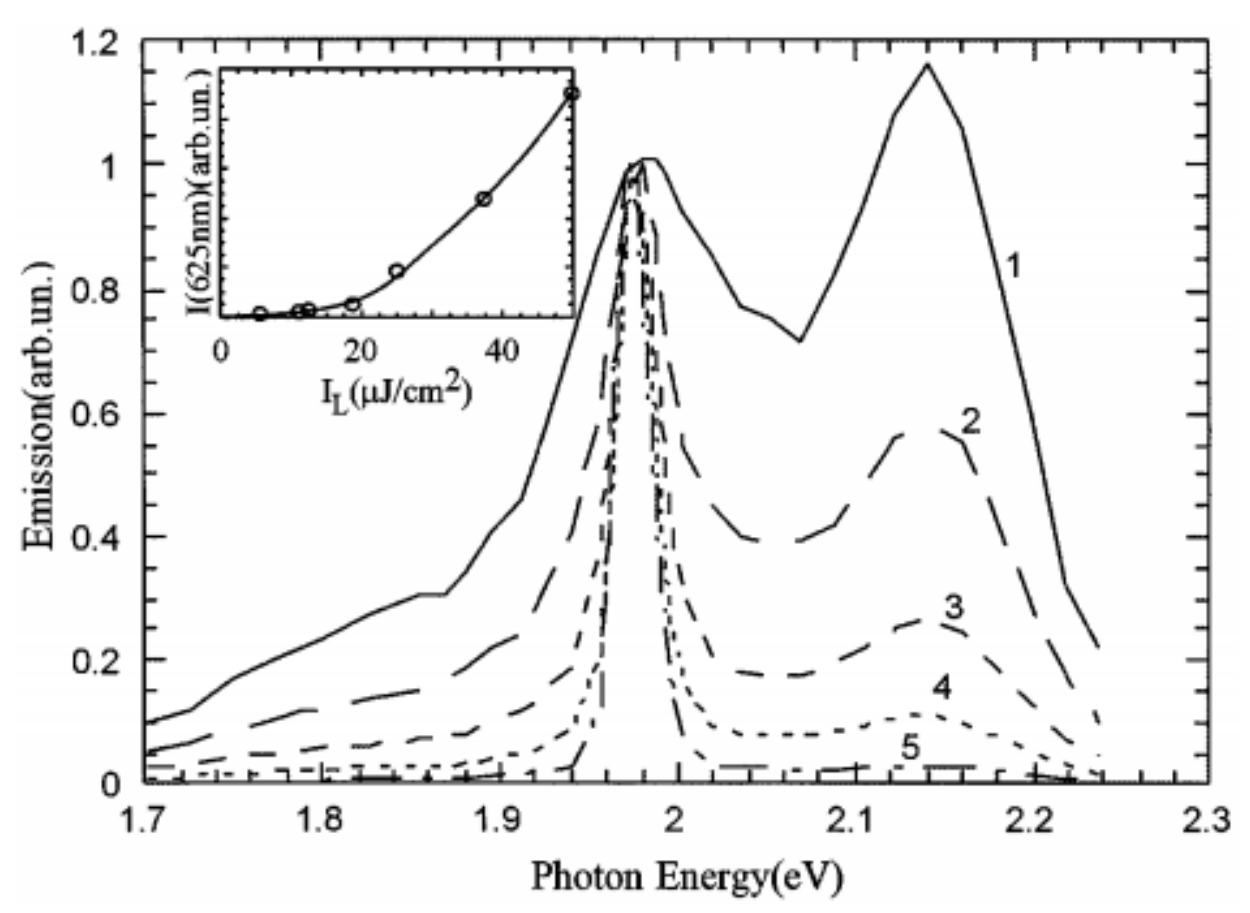}
\caption{Normalized emission spectra for a thin film of $\pi$-conjugated PPV derivatives  at various pulsed excitation fluences~\cite{FrolovetAl97PRL}. $I_{1}$ = 10\,$\mu$J/cm$^2$, $I_{2}$ = 2$I_{1}$ ($\times$ 1/3), $I_{3}$ = 3$I_{1}$ ($\times$ 1/8), I$_{4}$ = 5.4$I_{1}$ ($\times$ 1/26), $I_{5}$ = 25$I_{1}$ ($\times$ 1/200). The inset shows the amplification at 625\,nm close to the threshold intensity $I_{2}$. Reproduced with permission from~\cite{FrolovetAl97PRL}. Copyright 1997, American Physical Society.}
\label{SFPPV-1}
\end{figure}

Frolov {\it et al}.\ studied thin films of $\pi$-conjugated poly (p-phenylene vinylene) (PPV) derivatives at room temperature~\cite{FrolovetAl97PRL}. At low pump densities, a broad peak (with FWHM $\sim$ 80\,nm) was observed, but it collapsed into a much narrower (FWHM $\sim$ 7\,nm) and stronger emission peak at high densities ($n > n_{0} \sim$ 10$^{17}$\,cm$^{-3}$), accompanied by nonlinear amplification, as shown in Fig.\,\ref{SFPPV-1}. 
%
In an organic quaterthiophene semiconductor whose molecules are arranged in {\it H}-aggregate fashion, time-resolved photoluminescence measurements showed that the radiative lifetime had a linear correlation with the inverse of the number of the coherently emitting dipoles, i.e., $\tau_\textrm{rad} \propto 1/N$~\cite{MeinardietAl03PRL}. More recently, phosphorescence SR was demonstrated in heavy-metal-containing $\pi$-conjugated polymers, as a result of the significant spin-orbit interaction provided by the large atomic number elements~\cite{KhachatryanetAl12PRB}.  These experiments on cooperative emission from polymers and molecular crystals have also been discussed using the theory of SR for Frenkel excitons~\cite{TokihiroetAl93PRB}.

\subsection{Semiconductor Quantum Dots and Nanocrystals}

\begin{figure}[hbtp]
\centering
\includegraphics[scale=1]{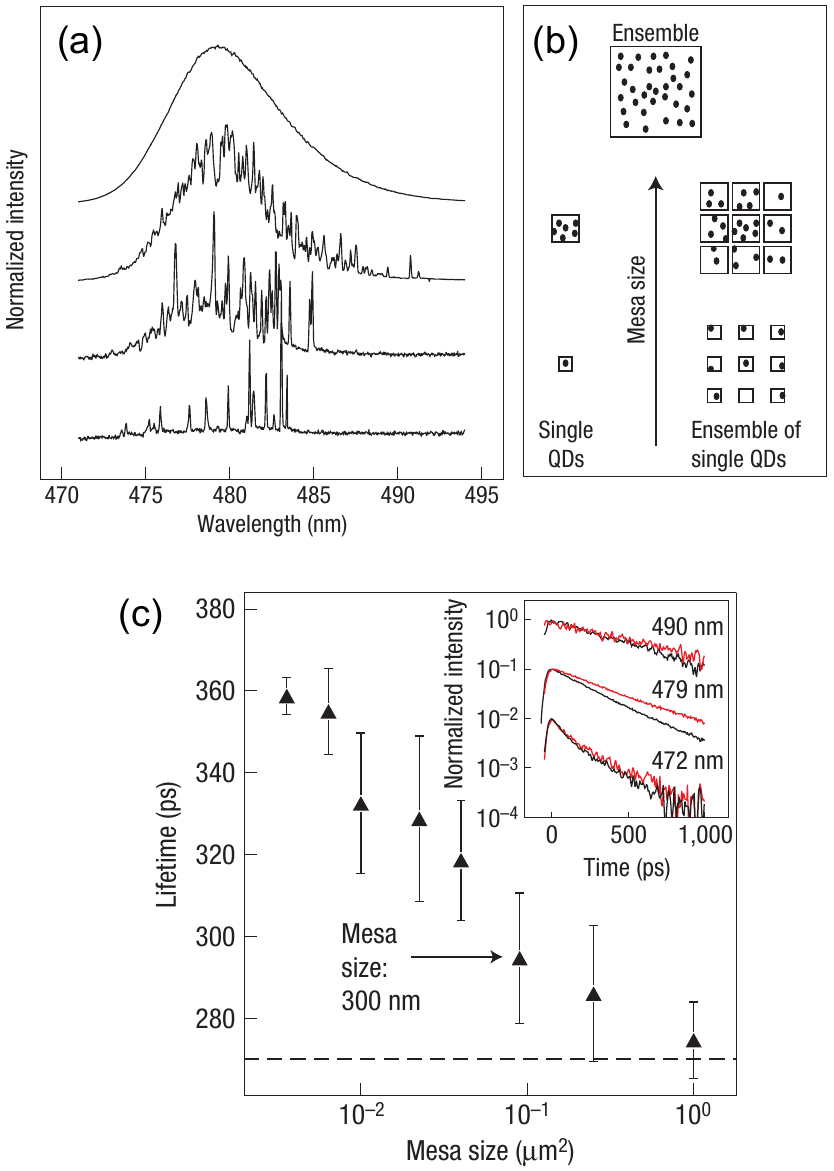}
\caption{Radiative coupling between self-assembled CdSe/ZnSe quantum dots as evidenced by a decay rate that increases with increasing number of dots within the wavelength.  (a)~Photoluminescence spectra of single mesas with edge sizes of 25\,$\mu$m, 1\,$\mu$m, 350\,nm, and 175\,nm (top to bottom). (b)~Schematic representations of a series of mesa-shaped samples containing different number of dots. (c)~Measured radiative decay times for mesas with different sizes, showing an accelerated decay rate for larger number of dots. Reproduced (adapted) with permission from~\cite{ScheibneretAL07NP}. Copyright 2007, Nature Publishing Group.}
\label{SRQDs}
\end{figure}

Zero-dimensional semiconductors, or artificial atoms, including quantum dots (QDs) and nanocrystals, provide another class of atomic-like systems in a solid-state environment for exploring cooperative emission~\cite{Schmitt-RinketAl87PRB,Takagahara87PRB,Kayanuma88PRB,SpanoetAl90PRL,ChenetAl03PRL,YukalovYukalova10PRB,OkuyamaEto12JPCS,TighineanuetAl15arXiv}.
Scheibner {\it et al}.~\cite{ScheibneretAL07NP} investigated light emission properties of self-assembled CdSe/ZnSe QDs with individual dot sizes of 6-10\,nm. In order to examine whether there exists any correlation between the number of QDs in the sample and the decay rate, they prepared a series of mesa-shaped samples with different sizes [see Figs.\,\ref{SRQDs}(a) and (b)] and measured photoluminescence lifetimes under weak excitation. They found that the decay rate increases with increasing mesa size [Fig.\,\ref{SRQDs}(c)], from which the average range of interaction between QDs was estimated to be $\sim$150\,nm; this value is much larger than the size of individual QDs but close to the effective wavelength of the emitted radiation for ZnSe ($\sim$180\,nm). Based on these observations, the authors claimed that superradiant QD-QD coupling occurs when one QD is placed within an average distance of one wavelength from another QD.

\begin{figure}[h]
\centering
\includegraphics[scale=0.81]{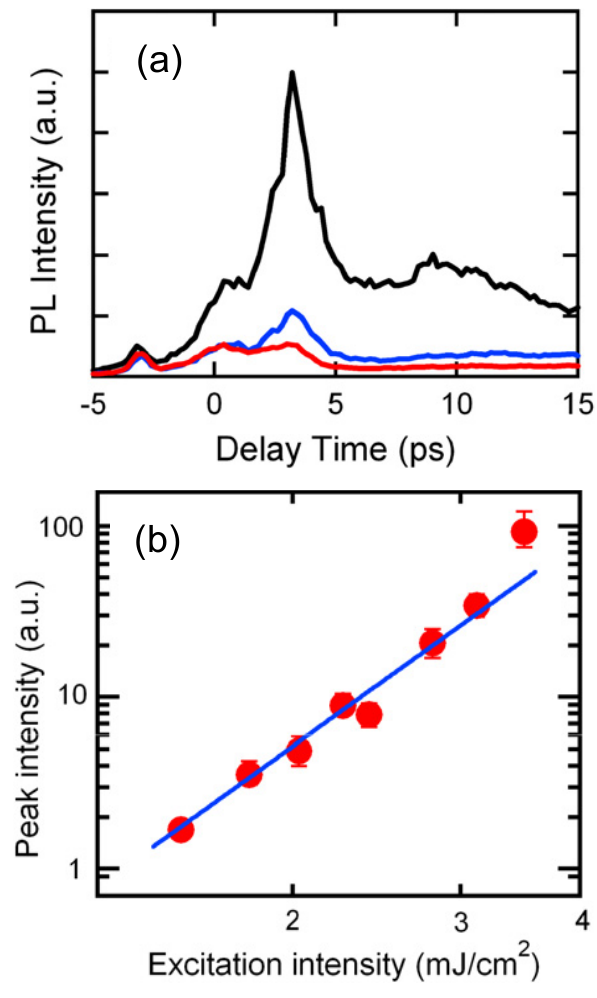}
\caption{Time-resolved evidence for cooperative emission from CuCl nanocrystals in a NaCl matrix.  (a)~Time-resolved photoluminescence signal with excitation intensities of 3.5\,mJ/cm$^2$ (black line), 2.80\,mJ/cm$^2$ (blue line), and 2.3\,mJ/cm$^2$ (red line). (b)~Photoluminescence peak intensity versus excitation intensity, showing superlinear dependence. The blue line shows the ideal superfluorescence behavior under the assumption that the number of excited dots is proportional to the square of the excitation intensity. Reproduced (adapted) with permission from~\cite{MiyajimaetAl09JPCM}. Copyright 2007, IOP Publishing.}
\label{QD-SF}
\end{figure}

More recently, time-resolved photoluminescence experiments on ensembles of CuCl nanocrystals embedded in a NaCl matrix have exhibited signs of cooperative emission, including a delayed pulse~\cite{MiyajimaetAl09JPCM,MiyajimaetAl11PSSC,PhuongetAl13JL}.  Population inversion between the biexciton and exciton states was efficiently achieved via resonant two-photon excitation of biexcitons; this strategy avoids direct excitation of the exciton state since the biexciton energy is smaller than twice the exciton energy.  Time profiles of photoluminescence for different excitation intensities are shown in Fig.\,\ref{QD-SF}(a). A clear peak appeared at a time delay of $\sim$3\,ps for the highest pump fluence (3.5\,mJ/cm$^2$), and its peak intensity increased superlinearly with the excitation intensity, as shown in Fig.\,\ref{QD-SF}(b). The peak intensity exhibited a fourth-power dependence on the excitation intensity, indicating that the density of excited
QDs is proportional to the square of the excitation intensity under two-photon excitation.


\section{Cooperative Spontaneous Emission from Extended States in Solids}
\label{Extended}

Section \ref{Localized} described some of the initial observations of cooperative spontaneous emission processes in solids.  However, in those systems, emission arose from an ensemble of atomic-like states, and thus, the essential physics was identical to that of SR and SF in atomic and molecular gases (Section \ref{Gases}).  Isolated atomic-like emitters (molecules, nanocrystals, and quantum dots) were embedded in a {\em passive} matrix without any free carriers around, and none of the bona fide solid-state physics elements, such as ultrafast dephasing, hot carrier relaxation via phonon emission, and excitonic correlations, were important.  In this section, genuine solid-state SR and SF phenomena, explicitly involving {\em extended} states and strong Coulomb interactions, are reviewed.

\subsection{Excitonic Superradiance in Semiconductor Quantum Wells}
\label{xSR}

A large body of work has been devoted to {\em excitonic} SR in semiconductors, where accelerated electron-hole recombination occurs through cooperation. Very fast ($\sim$ a few ps) decay of photoluminescence from resonantly excited excitons in semiconductor quantum wells (QWs) has been observed in a number of experimental studies~\cite{FledmannetAl87PRL,DeveaudetAl91PRL,VledderetAl99JL}. Theoretical studies ensued~\cite{Hanamura88PRB,AndreanietAl91SSC,Knoester92PRL,Citrin93PRB,BjorketAl96JOSAB}, and the following physical picture has emerged.

When the surface of a high-quality semiconductor crystal is illuminated by a coherent laser pulse at an excitonic resonance, a coherent polariton mode is excited. The coherent polarization decays due to various dephasing processes and also due to radiation from the surfaces. This radiative decay can be rather weak in the case of a bulk sample and at liquid helium temperatures at which most experiments have been done.  The situation changes in QWs, where the excitons are excited in a 2D layer of thickness much smaller than the wavelength ($\lambda$) of light at the resonance frequency~\cite{LeeLee74PRB}. In this case, the excitonic polarization strongly couples to the photon modes outgoing from the surface, and thus, its decay can be dominated by radiative processes. Since the coherent polarization has been resonantly excited by an ultrashort laser pulse within a large macroscopic area, it then decays radiatively as a {\em giant dipole}, i.e., the decay is superradiant by definition. The largest area from which the excitonic polarization coherently decays into a given electromagnetic mode is simply the size of a transverse mode, i.e., $\sim \lambda^2$. Therefore, the maximum enhancement of the radiative decay scales as the number of excitons within this area, i.e., in proportion to $(\lambda/a_\text{B})^2$, where $a_\text{B}$ is the Bohr radius of the 1$s$ exciton state. If the excitons are localized by scattering or disorder to a length $L_\text{c} < \lambda$, the enhancement scales as $(L_\text{c}/a_\text{B})^2$.  It has also been suggested that Bose-Einstein condensation of excitons may increase the coherence area and the resulting decay rate~\cite{ButovFilin98PRB,DaiMonkman11PRB}. 

Furthermore, cooperative emission properties of semiconductor QW systems can be modified and enhanced through quantum engineering of electronic and photonic states by optimization of periodicity, thicknesses, and dimensionality. Inter-QW superradiant coupling can be induced and/or employed in multiple-QW periodic structures with Bragg resonances~\cite{HuebneretAl99PRL,AmmerlahnetAl00PRB,IkawaCho02PRB,IvchenkoetAl04PRB}, quasi-periodic Fibonacci multiple-QW structures~\cite{ChangetAl14NJP}, and quasi-periodic double-period QW structures~\cite{ChangetAl15OE}.  Moreover, theoretical studies of excitonic SR in quantum wires~\cite{IvanovHaug93PRL,ManabeetAl93PRB,ChenetAl01PRB} and quantum dots~\cite{Schmitt-RinketAl87PRB,Takagahara87PRB,Kayanuma88PRB,SpanoetAl90PRL,ChenetAl03PRL,YukalovYukalova10PRB,OkuyamaEto12JPCS} have provided additional predictions and incentives for experimental studies.

\begin{figure}[hbtp]
\centering
\includegraphics[scale=1]{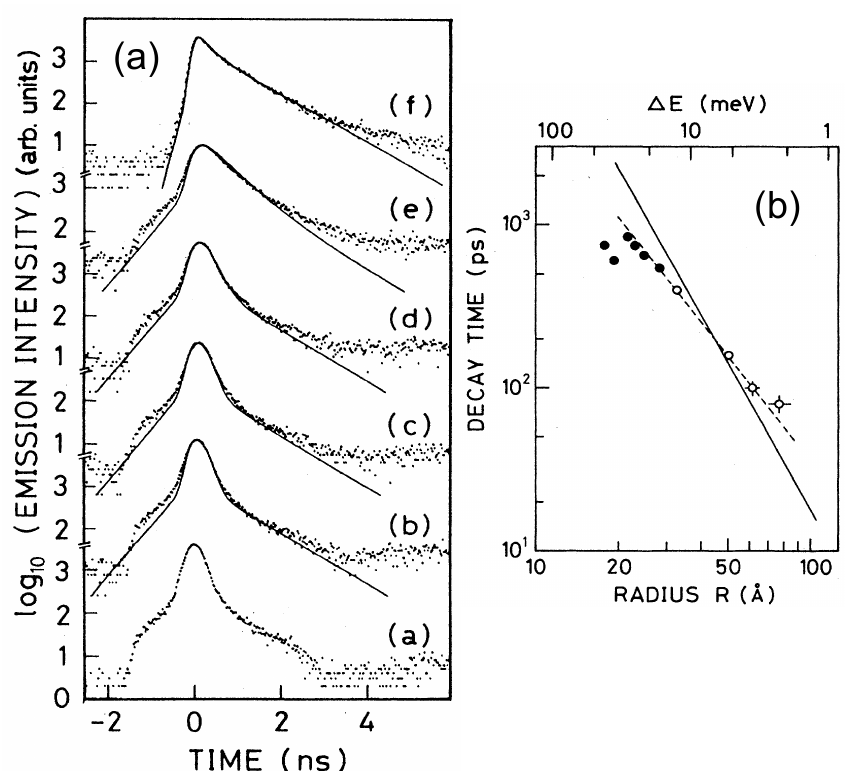}
\caption{(a)~Time-dependent photoluminescence data for CuCl nanocrystals with different radii, $R$.  Curve $a$: instrumental response function.   $b$:~$R =$ 77\,\AA.  $c$:~$R =$ 61\,\AA.  $d$: $R =$ 51\,\AA. $e$:~$R =$ 33\,\AA.  $f$:~$R =$ 25\,\AA.  (b)~Extracted decay time versus $R$.  Dashed line: $\propto$ $R^{2.1}$ dependence.  Solid line: calculated lifetime expected for excitonic SR. Reproduced (adapted) with permission from~\cite{NakamuraetAl89PRB}. Copyright 1989, American Physical Society.}
\label{QD-SR}
\end{figure}

Experimentally, size-dependent radiative decay in nanocrystals, expected for excitonic SR as described above, have been demonstrated~\cite{NakamuraetAl89PRB,ItohetAl90JL}.  Nakamura, Yamada, and Tokizaki studied the radiative decay of resonantly excited excitons confined in CuCl semiconducting nanocrystals with radii, $R$, of 18-77\,\AA \ in glass matrices. They found that the radiative decay
rate was proportional to $R^{2.1}$, which is consistent with a theoretical estimate based on excitonic SR~\cite{Hanamura88PRB}.  Figure~\ref{QD-SR}(a) shows time-resolved photoluminescence data for various nanocrystals with different sizes, while Fig.\,\ref{QD-SR}(b) plots the extracted decay time as a function of crystal size, together with a theoretical prediction (solid line).


\subsection{Superfluorescence from Semiconductor Quantum Wells}
\label{SF-QW}

As described in Sections \ref{Gases} and \ref{Localized} above, SF has been observed in many atomic and molecular systems since the 1970s. However, in semiconductor materials, SF has been difficult to observe due to the inherently fast scattering  of carriers. Typically, in semiconductors, photogenerated nonequilibrium carriers are spread over energy bands, limiting the number of dipole oscillators, e.g., electron-hole ($e$-$h$) pairs, within the radiation bandwidth, keeping the cooperative frequency below the threshold for achieving SF.  One possible way to overcome these limitations is to place the system in a strong perpendicular magnetic field ($B$) and at low temperature ($T$)~\cite{BelyaninetAl91SSC,BelyaninetAl97QSO}. A strong $B$ can effectively increase the dipole moment as well as the number of carriers contributing to SF, through wavefunction shrinkage and an increase in the density of states. Scattering is suppressed in a strong $B$ due to the reduced phase space available for scattering, which leads to longer relaxation times [effective $T_1$ and $T_2$ in Eq.\,(1)].  Low $T$ increases quantum degeneracy and suppress scattering.

By optically exciting $e$-$h$ pairs in an InGaAs/GaAs quantum well (QW) system, we have made the first SF observation using extended states in a solid~\cite{NoeetAl12NP}. This provides a good system in which to study many-body physics in a highly controllable environment through $B$, $T$, and pair density (laser power, $P$).  For an $e$-$h$ plasma in semiconductor QWs, quantum dipole oscillators are $e$-$h$ pairs, and their SF is a process of collective radiative recombination. A high enough pump fluence is needed to provide strong Fermi degeneracy of the photoexcited nonequilibrium carriers. This maximizes the population inversion in Eq.~(\ref{coop}) and ensures that stimulated recombination prevails over the inverse process of the interband absorption. In addition, Fermi degeneracy gives rise to a many-body Coulomb enhancement of the gain (Fermi edge singularity)~\cite{Schmitt-RinketAl86PRB,KimetAl13SR}, which makes coherent and cooperative spontaneous emission possible even without a strong $B$ if $T$ is sufficiently low.

\subsubsection{Sample and Experimental Methods}

\begin{figure}[hbtp]
\centering
\includegraphics[scale=0.39]{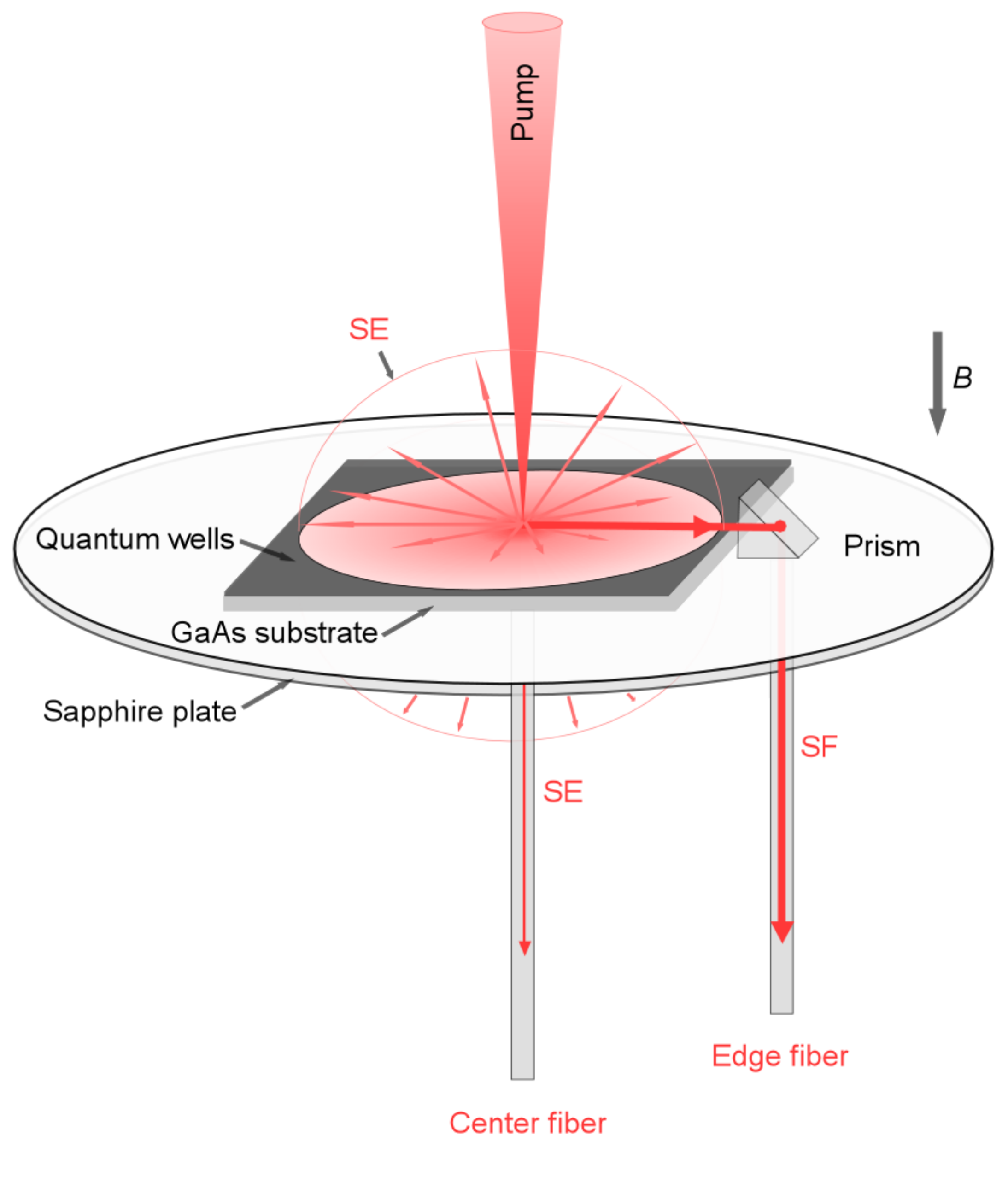}
\caption{Schematic diagram of the experimental geometry for observing coherent spontaneous emission from photoexcited semiconductor QWs in a magnetic field.  The magnetic field is applied perpendicular to the QWs, parallel to the incident pump beam.  Spontaneous emission (SE) is emitted isotropically, while SF is emitted in the QW plane and detected through the edge fiber. Reproduced (adapted) with permission from~\cite{CongetAl15PRB}. Copyright 2015, American Physical Society.}
\label{ExperimentalMethods}
\end{figure}

The sample we studied was a multiple-QW sample grown by molecular beam epitaxy, consisting of 15 layers of 8-nm In$_{0.2}$Ga$_{0.8}$As separated by 15-nm GaAs barriers grown on a GaAs buffer layer and GaAs (001) substrate. The confinement of the QW potential resulted in the formation of a series of subbands, both for electrons in the conduction band and holes in the valence band. Due to the strain caused by the lattice mismatch between the In$_{0.2}$Ga$_{0.8}$As and GaAs layers, a relatively large energy splitting (75\,meV) occurred between the $E_{1}H_{1}$ and $E_{1}L_{1}$ subbands, so only the $E_{1}H_{1}$ transition was relevant to our spectral range.  In the presence of an external $B$ applied perpendicular to the QW plane, each subband splits into a series of peaks due to Landau quantization. For example, the $E_{1}H_{1}$ transition splits into ($N_\text{e}$,$N_\text{h}$) = (00), (11), (22), ... transitions, where $N_\text{e}$ ($N_\text{h}$) is the electron (hole) Landau level (LL) index.

We performed time-integrated photoluminescence (TIPL) spectroscopy, time-resolved photoluminescence (TRPL) spectroscopy, and time-resolved pump-probe spectroscopy measurements on the InGaAs QW sample under a variety of $B$, $T$, and $P$ conditions, at the Ultrafast Optics Facility of the National High Magnetic Field Laboratory in Tallahassee, Florida (using either a 31-T DC resistive magnet or a 17.5-T superconducting magnet), and at Rice University with a 30-T pulsed magnet system~\cite{NoeetAl13RSI}. The main laser system used was an amplified Ti:sapphire laser (Clark-MXR, Inc., CPA 2001, or Coherent Inc., Legend), producing 150\,fs pulses of 775\,nm (1.6\,eV, with CPA 2001) or 800\,nm (1.55\,eV, with Legend) radiation at a repetition rate of 1\,kHz. In addition, an optical parametric amplifier (OPA) was used to produce intense outputs with tunable wavelengths between
850\,nm and 950\,nm.

For TIPL and TRPL measurements, the sample was mounted on a sapphire plate, and a $\mu$-prism was placed at one edge of the sample to redirect in-plane emission; see Fig.\,\ref{ExperimentalMethods}.  Two fibers, center and edge fibers, were used for PL collection; the former was used for monitoring spontaneous emission (which was emitted in all 4$\pi$ spatial directions with equal probability), while the latter was used to observe SF (which was emitted in the plane of the QWs)~\cite{NoeetAl12NP,KimetAl13PRB,KimetAl13SR,CongetAl15PRB}. TIPL was measured with a CCD-equipped monochromator, and TRPL was measured either using a streak camera system or a Kerr-gate method. Pump-probe measurements were made in a transmission geometry in the Faraday configuration, where the pump and probe beams were parallel to the $B$ and normal incident to the QWs. For a particular transition, the differential transmission, $\Delta T/T$, was monitored by a photodiode, which is proportional to the population inversion for that transition.

\subsubsection{Time-Integrated Emission Spectra}

\begin{figure}[h]
\centering
\includegraphics[scale=0.75]{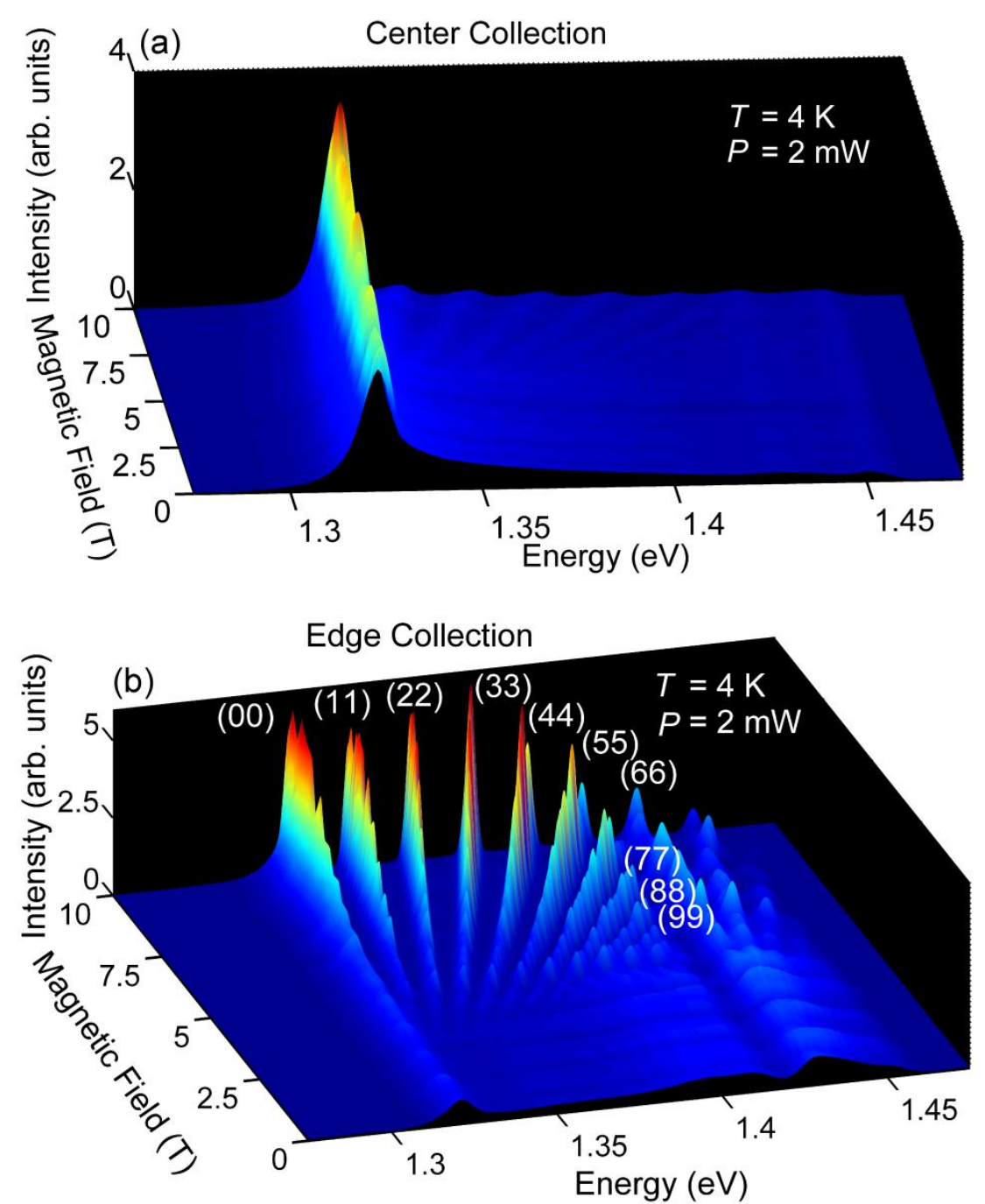}
\caption{Magnetic field dependence of time-integrated PL collected with the (a) center fiber and (b) edge fiber
at 4\,K with an average excitation laser power of 2\,mW. Reproduced with permission from~\cite{CongetAl15PRB}. Copyright 2015, American Physical Society.}
\label{TIPL_B}
\end{figure}

TIPL showed very different behaviors between the center and edge collections. Figures~\ref{TIPL_B}(a) and \ref{TIPL_B}(b) show the $B$ dependence of time-integrated center-fiber and edge-fiber-collected PL, respectively, from $B =$ 0\,T to 10\,T with $P =$ 2\,mW and at $T =$ 4\,K~\cite{CongetAl15PRB}. The only feature observed in the center-fiber-collected PL, shown in Fig.\,\ref{TIPL_B}(a), is the lowest-energy, ($N_\text{e}$,$N_\text{h}$) $=$ (00) transition, whose emission peak slightly blue-shifts with increasing $B$ through the diamagnetic shift~\cite{AkimotoHasegawa67JPSJ}. In contrast, the intensity of edge PL emission drastically increases with $B$, as shown in Fig.\,\ref{TIPL_B}(b). In the low $B$ regime ($<$~4\,T), the edge PL emission spectrum is characterized by two peaks at $\sim$1.32\,eV and $\sim$1.43\,eV, corresponding to the $E_{1}H_{1}$ 1$s$ and $E_{1}L_{1}$ 1$s$ transitions, respectively, whose shape and intensity are more or less stable with $B$. However, a further increasing in $B$ leads to emission from other LLs, which becomes much brighter, sharper, and better spectrally separated. As shown in Fig.\,\ref{TIPL_B}(b), ten peaks, due to the (00), (11), ... (99) interband transitions, can be clearly observed. These differences of PL emission between the center and edge collections lie in the gain distribution in the InGaAs QW system. Optical gain exists only for electromagnetic waves propagating along the QW plane, which leads to in-plane SF emission; no optical gain is available in the direction perpendicular to the QW plane, leading to ordinary spontaneous emission (SE) in the center collection.

\subsubsection{SF Bursts in Strong Magnetic Fields}

\begin{figure}[b]
\centering
\includegraphics[scale = 0.52]{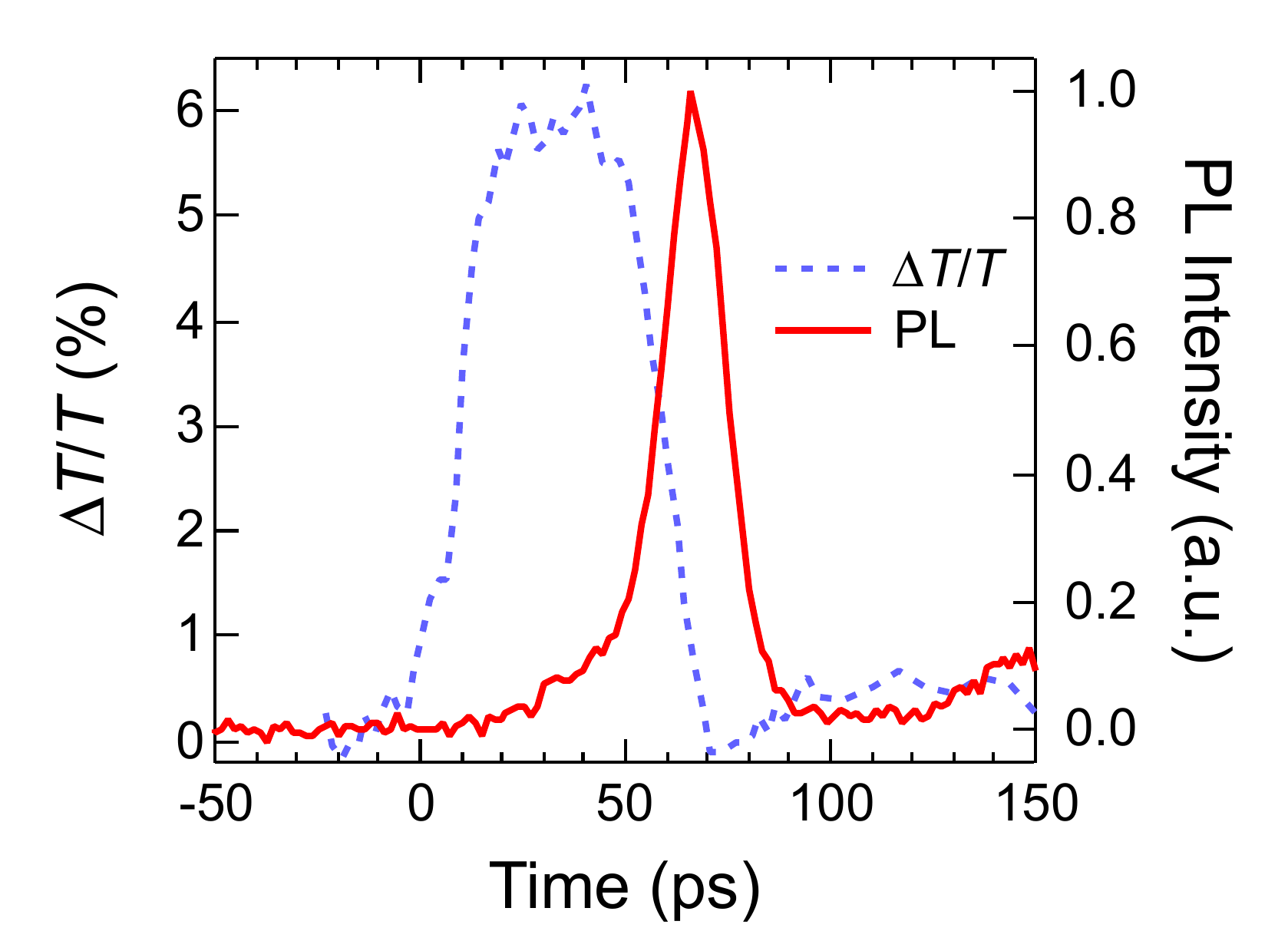}
\caption{Simultaneously taken pump-probe and TRPL data for the (22) transition at 17\,T and 5\,K, showing direct evidence of SF in the InGaAs QW system.}
\label{SF_evidence}
\end{figure}

In order to obtain the direct evidence of SF emission, time-resolved pump-probe and PL measurements were performed at the same time. Figure \ref{SF_evidence} shows the simultaneously taken pump-probe differential transmission and TRPL data for the (22) transition at 17\,T and 5\,K~\cite{NoeetAl12NP,NoeetAl13FP}. Here, the differential transmission corresponds to the population dynamics in the system. After the optical pump, the population inversion is quickly built up in the system, then suddenly drops to zero at a delay time around 70\,ps, while at the same time a strong pulse of emission appears, as indicated by the TRPL data. Generally, in order to observe this salient SF emission feature, a high $B$, low $T$, and large $P$ is required.

\begin{figure*}[htbp]
\centering
\includegraphics[width = 1 \linewidth]{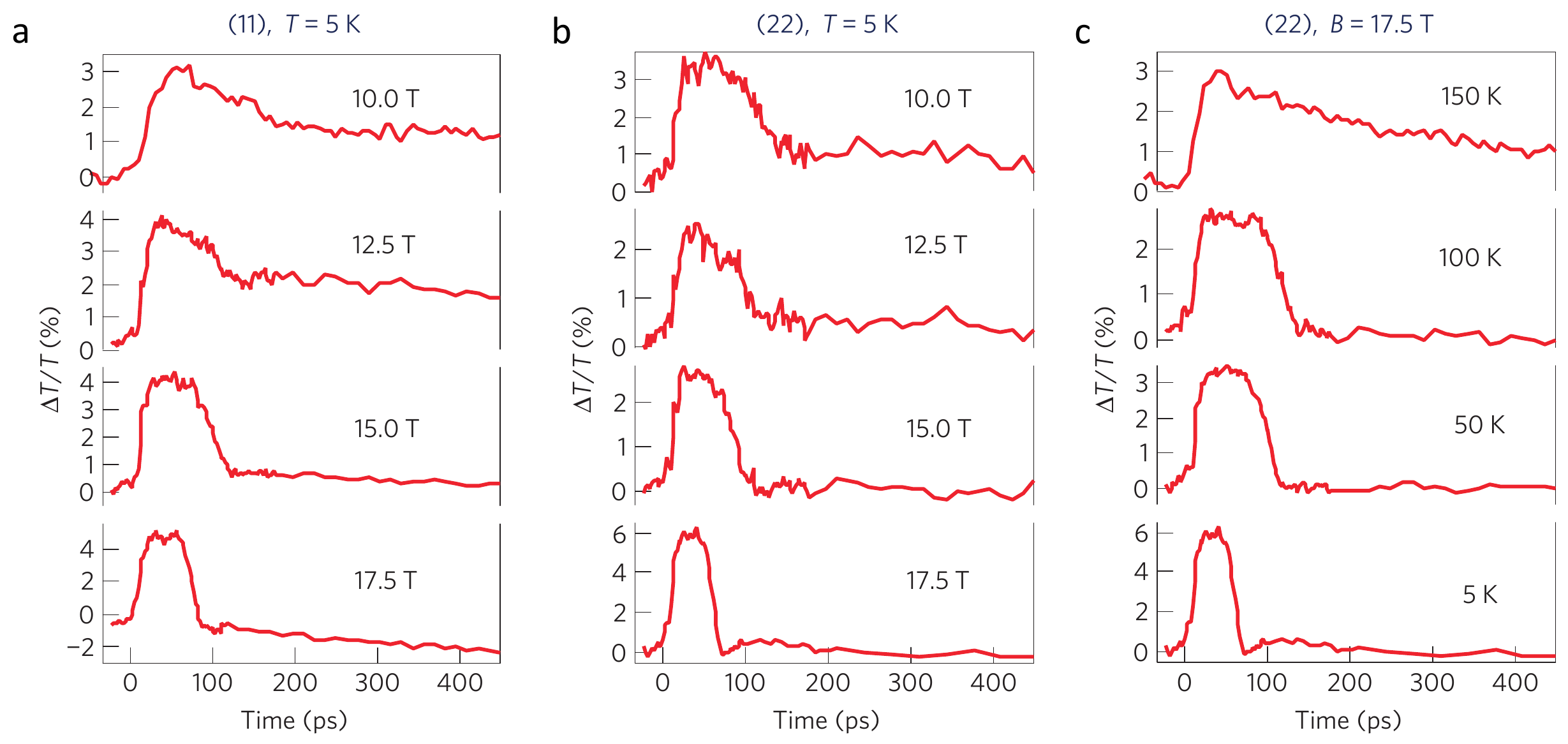}
\caption{Magnetic-field-dependent time-resolved pump-probe differential transmission traces probing the (a)~(11) and (b)~(22) transitions at 5\,K. (c)~Temperature-dependent pump-probe differential transmission traces probing the (22) transition at 17.5\,T. Reproduced (adapted) with permission from~\cite{NoeetAl12NP}. Copyright 2012, Nature Publishing Group.}
\label{SF_pumpprobe}
\end{figure*}

Results of pump-probe measurements on the (00) and (11) transitions at different $B$ are summarized in Fig.\,\ref{SF_pumpprobe}~\cite{NoeetAl12NP,NoeetAl13FP}. Figure~\ref{SF_pumpprobe}(a) shows $B$-dependent time-resolved pump-probe differential transmission for the (11) transition at 5\,K. At low $B$, such as 10\,T, the population difference of the (11) transition exhibits a slow exponential decay.  However, as $B$ increases, the temporal profile begins to exhibit a sudden drop that, with increasing $B$, becomes faster and sharper and occurs at a shorter time delay, becoming $\sim$80\,ps at 17.5\,T. The (22) transition shows a similar dependence on $B$, except that the population drops at an even earlier delay time compared with that of the (11) transition ($\sim$60\,ps at 17.5\,T), as shown in Fig.\,\ref{SF_pumpprobe}(b). Figure~\ref{SF_pumpprobe}(c) shows that a decreasing $T$ has a similar effect to an increasing $B$, which leads to a more sudden decrease in population that occurs at a shorter delay time as $T$ changes from 150\,K to 5\,K.

\begin{figure}[htbp]
\centering
\includegraphics[width=\linewidth]{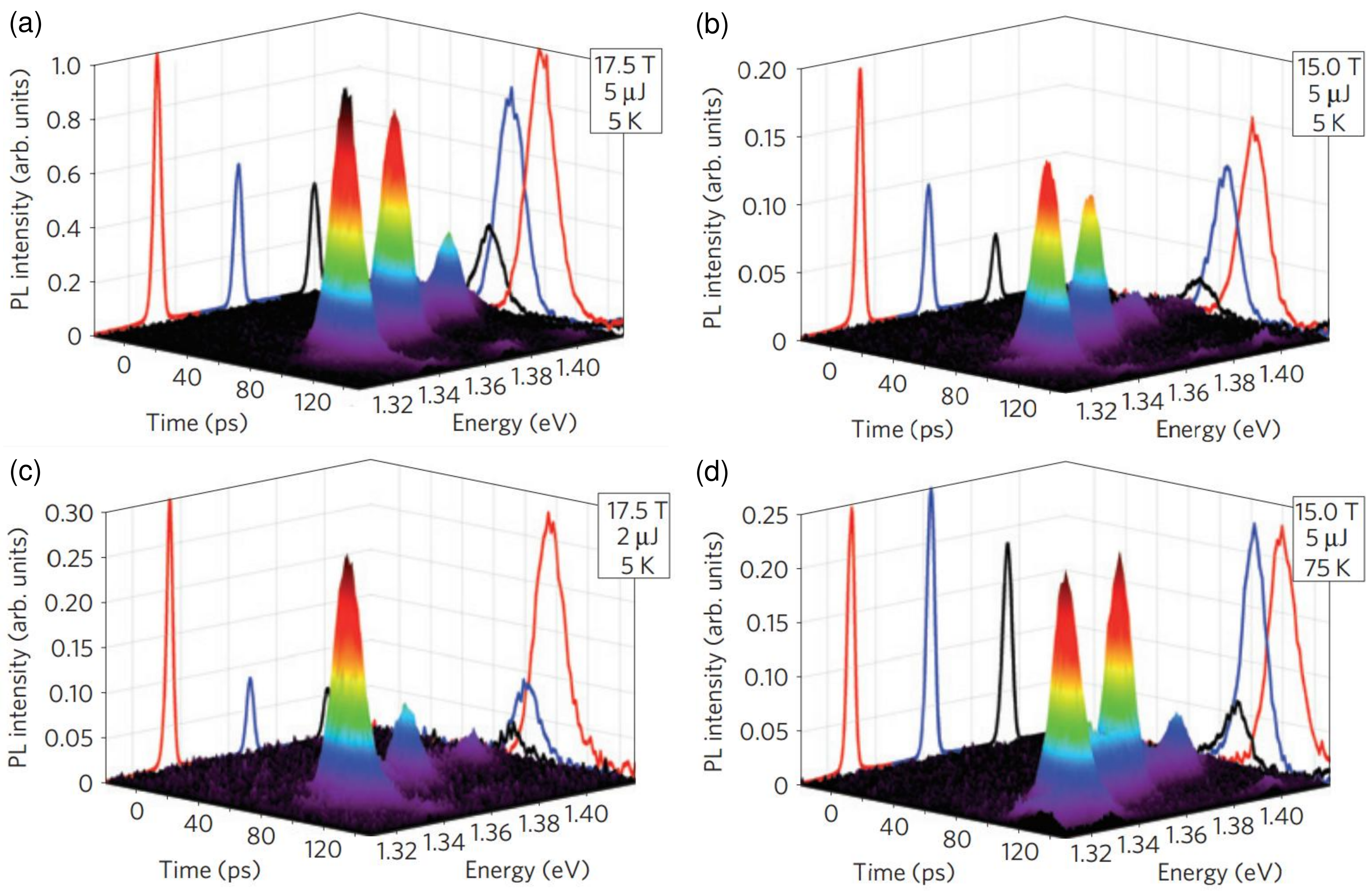}
\caption{Steak camera images of SF bursts as a function of photon energy and delay time at different magnetic fields, temperatures and pump pulse energies. (a)~17.5\,T, 5\,K, and 5\,$\mu$J. (b)~15\,T, 5\,K, and 5\,$\mu$J. (c)~17.5\,T, 5\,K, and 2\,$\mu$J. (d)~15\,T, 75\,K, and 5\,$\mu$J. Reproduced (adapted) with permission from~\cite{NoeetAl13FP}. Copyright 2012, Wiley.}
\label{SF_TRPL}
\end{figure}

We measured spectrally and temporally resolved SF bursts at different $B$, $T$, and pump pulse energies, as shown in Fig.\,\ref{SF_TRPL}. Figure~\ref{SF_TRPL}(a) shows a PL intensity map as a function of time delay and photon energy at 17.5\,T, 5\,K, and 5\,$\mu$J~\cite{NoeetAl12NP,NoeetAl13FP}. Three SF bursts, coming from the (00), (11), and (22) transitions, are clearly resolved, both in time and energy. Each burst emerges after a time delay, and the delay is longer for lower LLs, i.e., the highest-energy transition, (22), emits a pulse first, and each lower-energy transition emits a pulse directly after the transition just above it. Figure~\ref{SF_TRPL}(b) shows the effects of a reduced $B$ on SF bursts: smaller energy separations between LL, and a longer delay time for a given transition. Figure~\ref{SF_TRPL}(c) suggests that a lower pump pulse energy leads to weaker SF emission. Increasing $T$ has a similar effect on SF emission to a decreasing $B$, as shown in Fig.\,\ref{SF_TRPL}(d). With increasing $T$, emission from all transitions weakens significantly and moves to later delay times.

\begin{figure}[hbtp]
\centering
\includegraphics[scale=0.85]{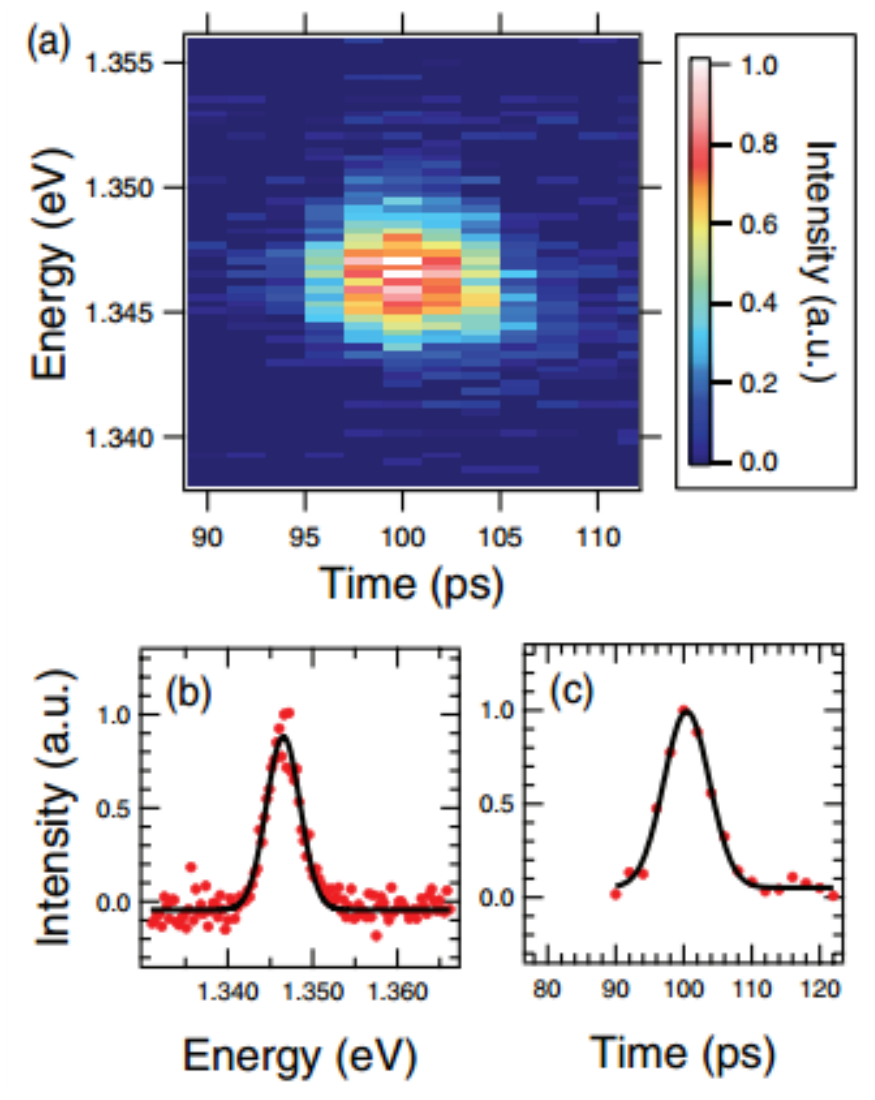}
\caption{(a)~SF burst for the (11) transition at 10\,T and 19\,K measured with a Kerr-gate method. (b)~Vertical slice at the intensity peak, showing a spectral width of $\sim$5\,meV. (c)~Horizontal slice at the intensity peak, showing a pulse width of $\sim$10\,ps. Reproduced (adapted) with permission from~\cite{NoeetAl13RSI}. Copyright 2013, AIP Publishing.}
\label{SF_Kerr}
\end{figure}

Due to a relatively large dispersion due to the graded-index collection fiber used and the monochromator in front of the streak camera, the time resolution in the above SF studies was limited to 20--30\,ps, which is not high enough to measure the true pulse widths of the SF bursts. In order to provide information on the widths of SF bursts quantitatively, TRPL measurements via a Kerr-gate technique were performed with a 30-T pulsed magnet in free space. Figure~\ref{SF_Kerr} shows a SF burst for the (11) transition at 10\,T and 19\,K~\cite{NoeetAl13RSI}. By taking vertical and horizontal slices at the peak of the burst, the pulse width was estimated to be $\sim$ 10\,ps, and the spectral width to be $\sim$ 5\,meV. Further investigation is needed to determine  how the pulse width and delay time vary with $B$, $T$, and $P$.

\subsubsection{Fluctuations in SF Pulse Direction}

\begin{figure}[hbtp]
\centering
\includegraphics[width=\linewidth]{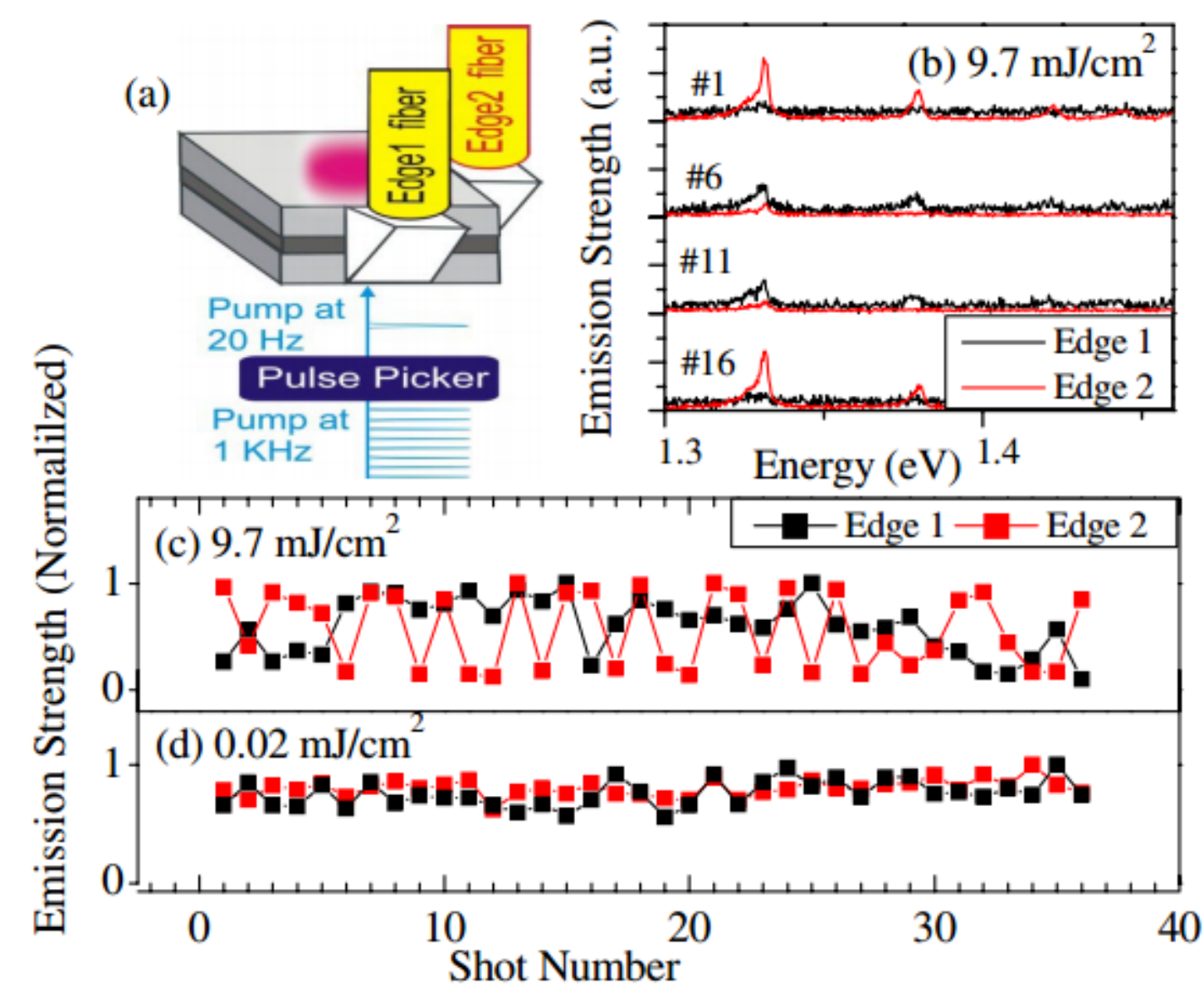}
\caption{(a)~Schematic setup for single-shot TIPL measurements. (b)~Four representative TIPL emission spectra from two edges at high fluence. Normalized emission strength for the (00) transition versus shot number in the (c)~SF regime and (d)~ASE or regime. Reproduced (adapted) with permission from~\cite{JhoetAl06PRL}. Copyright 2006, American Physical Society.}
\label{SF_direction}
\end{figure}

As described in Section \ref{intro}, randomness is expected in SF properties, such as intensities, pulse widths, delay times, and directions, due to quantum fluctuations~\cite{HaakeetAl79PRL,VrehenetAl80Nature,RehlerEberly71PRA,BonifacioLugiato75PRA,BonifacioLugiato75PRA2}. In order to investigate randomness in the direction of SF emission in the present case, single-shot TIPL measurements were performed in a two-fiber geometry, as indicated in Fig.\,\ref{SF_direction}(a), where emissions from two edge fibers were simultaneously taken upon single pulse excitation~\cite{JhoetAl06PRL,JhoetAl10PRB}. The measurement was taken under two excitation conditions: 9.7\,mJ/cm$^{2}$ with a 0.5-mm spot size (corresponding to the SF regime), and 0.02\.mJ/cm$^{2}$ with a 3-mm spot size (corresponding to the ASE or SE regime). Some representative spectra for the (00) transition at 9.7\,mJ/cm$^{2}$ is shown in Fig.\,\ref{SF_direction}(b). Figure~\ref{SF_direction}(c) plots the normalized emission strength for the (00) peak versus shot number under the high (9.7\,mJ/cm$^{2}$) and low (0.02\,mJ/cm$^{2}$) excitation, from which a strong anti-correlation signal from the two fibers can be observed at high pump fluence, indicating a collimated but a randomly changing SF emission direction from shot to shot. In Fig.\,\ref{SF_direction}(d), omnidirectional emission on every shot is observed, as expected in the ASE or SE regime.

\subsubsection{Many-Body Coulomb Enhancement of SF at the Fermi Edge}


\begin{figure*}[htbp]
\centering
\includegraphics[width=\linewidth]{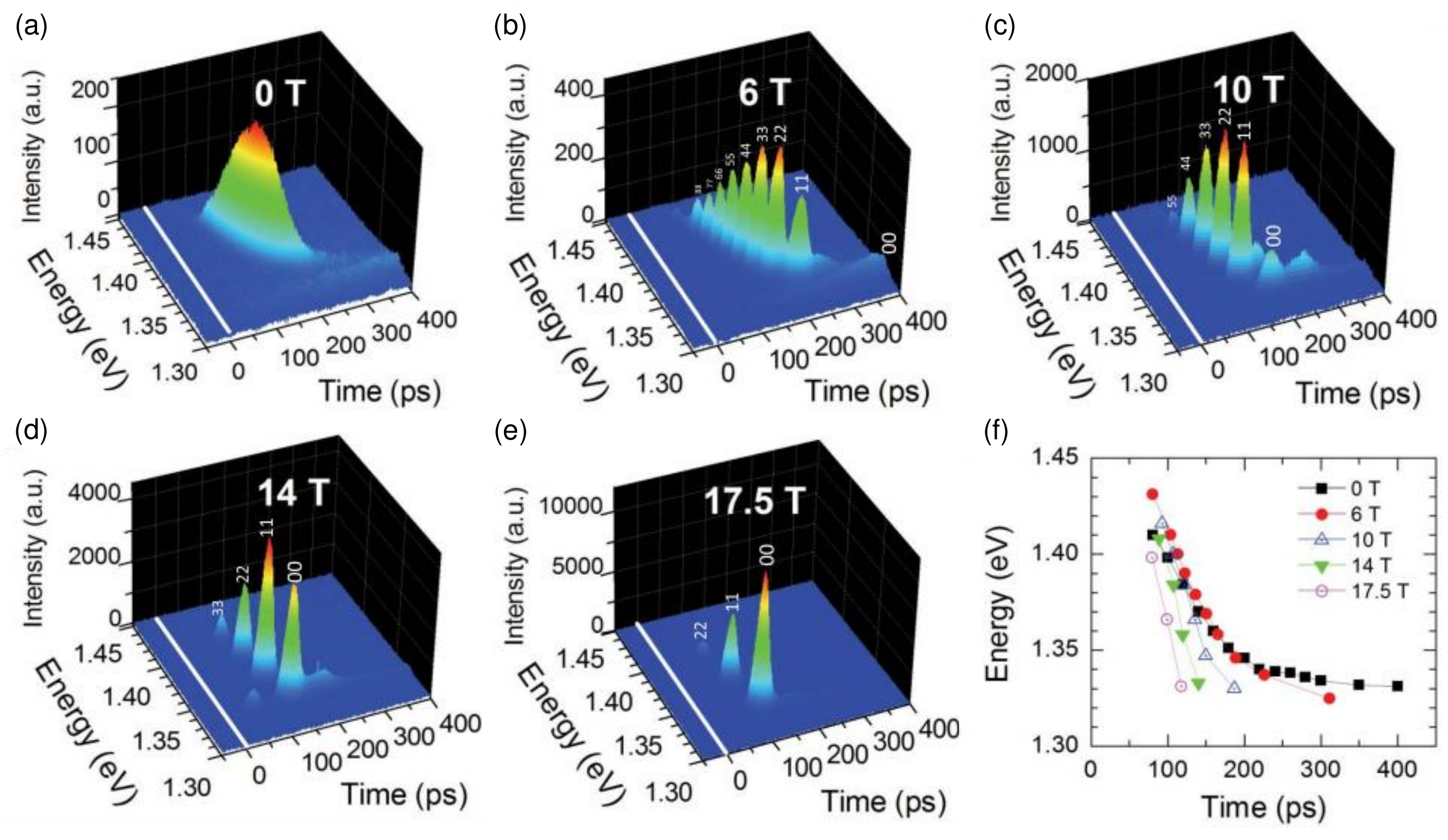}
\caption{Time-resolved PL spectra at (a)~0\,T, (b)~6\,T, (c)~10\,T, (d)~14\,T, (e)~17.5\,T with an excitation power of  2\,mW at 5\,K. (f)~Peak shift of emission as a function of time at different magnetic fields. Reproduced (adapted) with permission from~\cite{KimetAl13SR}. Copyright 2013, Nature Publishing Group.}
\label{TRPL_B}
\end{figure*}

As shown in Fig.\,\ref{SF_TRPL}, SF bursts from different LL transitions occur in a sequential manner: SF from the highest occupied LL is emitted first, which is followed by emission from lower and lower LLs. When the magnetic field or temperature is changed, the delay time for each SF burst changes; however, the delay times of different bursts change in such a way that the sequential order is preserved.  Namely, the relative timing of the bursts coming from different LLs is not random.

To further this sequential emission process, TRPL measurements were performed using a streak camera at different $B$ while $T$ and $P$ were kept constant.  Figures~\ref{TRPL_B}(a)-\ref{TRPL_B}(e) show SF emission as a function of photon energy and time delay at various $B$~\cite{KimetAl13SR}. With increasing $B$, the number of peaks decreases, and the energy separation between LLs increases due to increasing Landau quantization. Interestingly, at low $B$, the emission is characterized by a red-shifting continuum, which gradually evolves into discrete SF bursts at high $B$. At a given $B$, sequential SF emission is clearly observed: SF emission occurs only after all higher-energy SF bursts occur, and the delay time is longer for a burst from a lower LL. This sequential behavior can be even more clearly seen in Fig.\,\ref{TRPL_B}(f), which summarizes the peak positions of the SF bursts as a function of photon energy and time. Furthermore, Fig.\,\ref{TRPL_B}(f) indicates that, at the same photon energy, a higher $B$ can induce a SF burst earlier.

\begin{figure*}[hbtp]
\centering
\includegraphics[width=\linewidth]{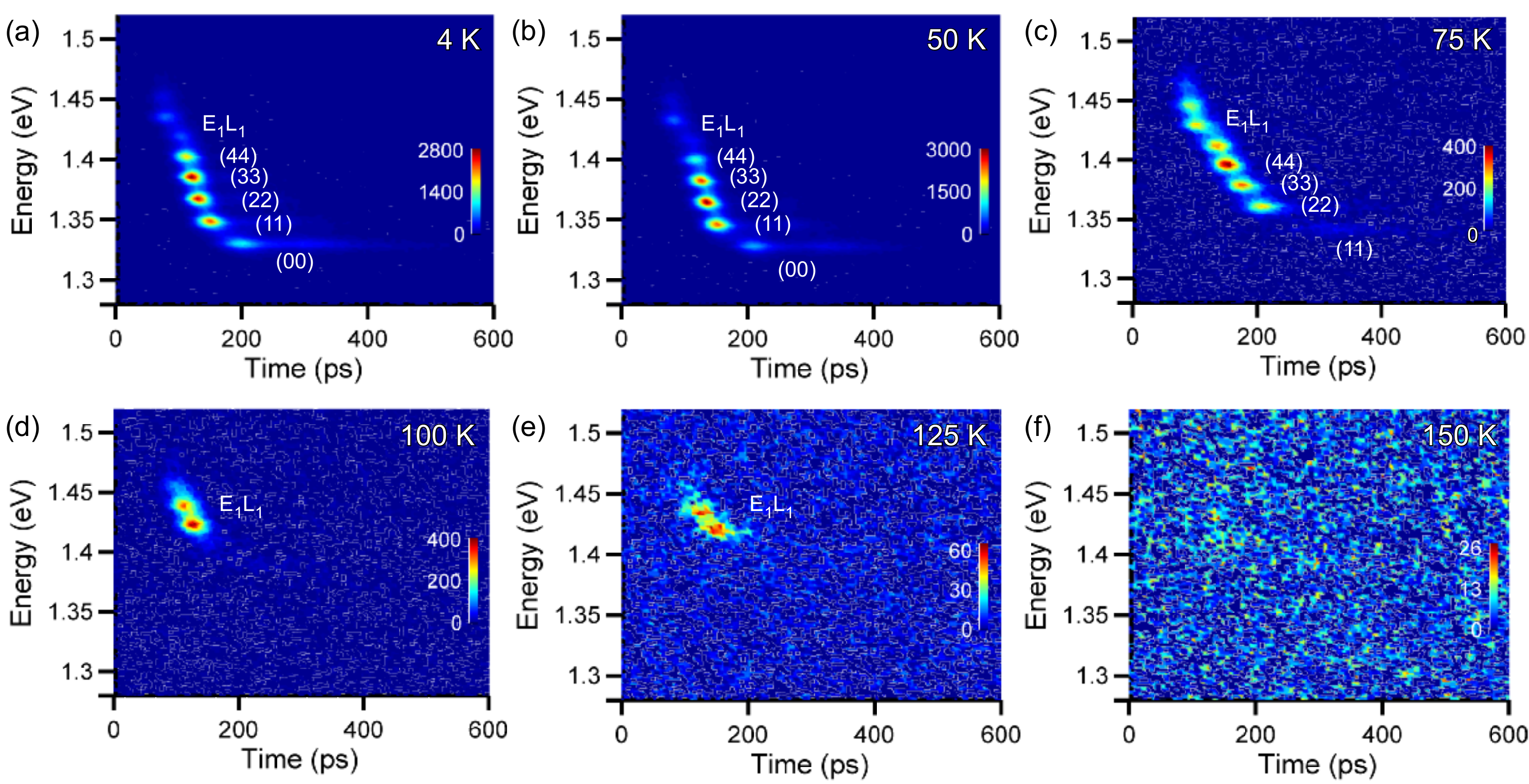}
\caption{Time-resolved PL spectra at different temperatures at a magnetic field of 10\,T and an excitation laser power of 2\,mW. Reproduced (adapted) with permission from~\cite{CongetAl15PRB}. Copyright 2015, American Physical Society.}
\label{TRPL_T}
\end{figure*}

\begin{figure}[hbtp]
\centering
\includegraphics[scale = 0.51]{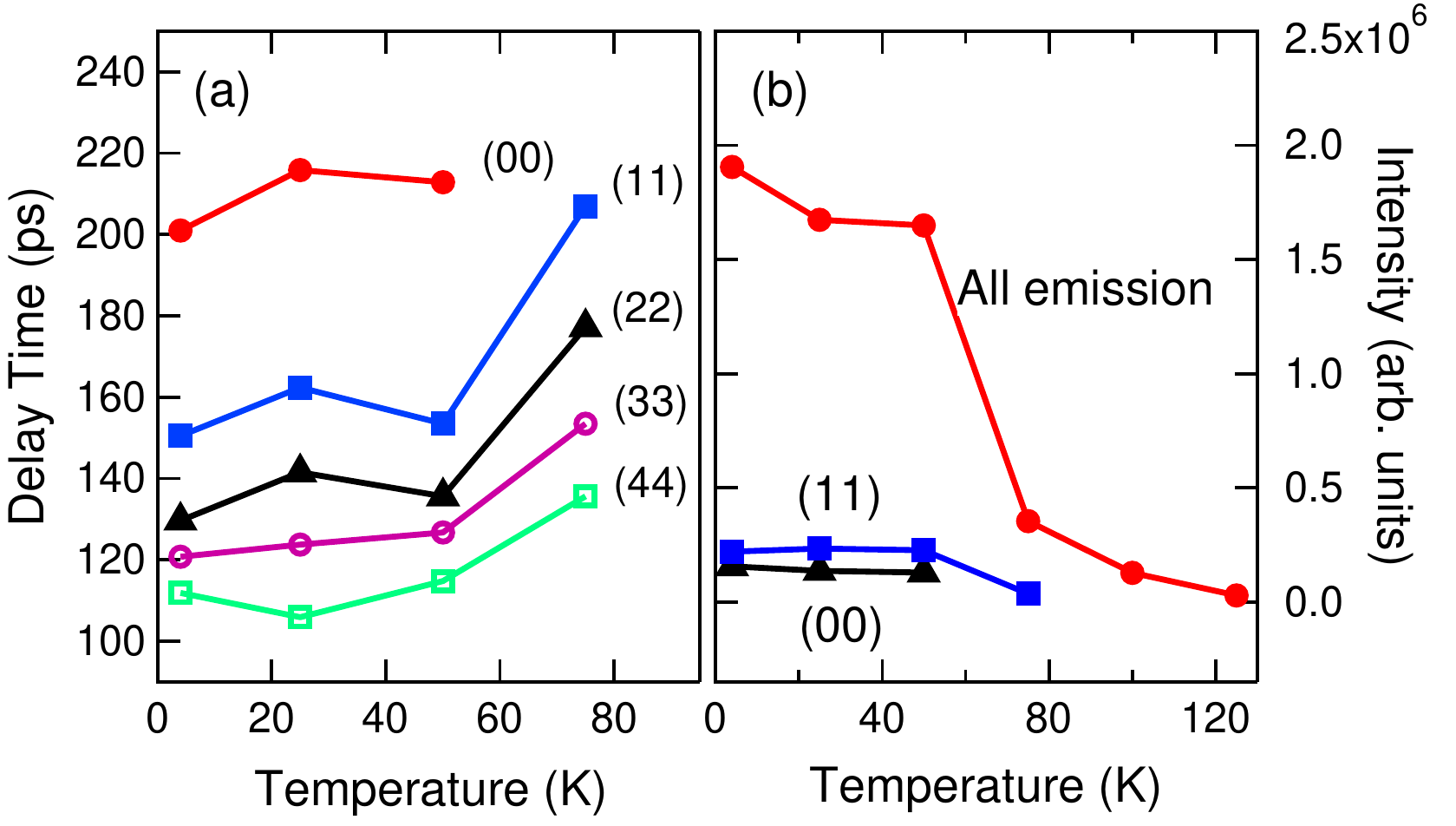}
\caption{(a)~Temperature dependence of SF delay times for different transitions at 10\,T and 2\,mW. (b)~Temperature dependence of spectrally and temporally integrated SF peak intensities at 10\,T and 2\,mW for all emission peaks as well as the (00) and (11) peaks. Reproduced (adapted) with permission from~\cite{CongetAl15PRB}. Copyright 2015, American Physical Society.}
\label{TRPL_T_DI}
\end{figure}

Figures \ref{TRPL_T}(a)-\ref{TRPL_T}(f) demonstrate that increasing $T$ has an effect similar to decreasing $B$ on the delay time of SF emission~\cite{CongetAl15PRB}. These data were taken at 10\,T and 2\,mW at $T =$ (a)~4\,K, (b)~50\,K, (c)~75\,K, (d)~100\,K, (e)~125\,K, and (f)~150\,K. At each $T$, multiple SF bursts coming from different LLs can be seen, with delay times that are shorter for those arising from higher LLs. With increasing $T$, the intensity of SF gradually decreases, and finally, at $T$ $>$ 150\,K, no SF bursts can be observed. Figure~\ref{TRPL_T_DI}(a) plots the temperature dependence of SF delay times for different transitions. The delay time increases monotonically with increasing $T$ for all peaks except the (00) peak. Figure~\ref{TRPL_T_DI}(b) shows the temperature dependence of integrated intensities for different transitions, showing that SF vanishes at high $T$.

We interpret these phenomena in terms of Coulomb enhancement of gain near the Fermi energy in a high-density $e$-$h$ system, which results in a preferential SF burst near the Fermi edge.  After relaxation and thermalization, the photogenerated carriers form degenerate Fermi gases with respective quasi-Fermi energies inside the conduction and valence bands. The recombination gain for the $e$-$h$ states just below the quasi-Fermi energies is predicted to be enhanced due to Coulomb interactions among carriers~\cite{Schmitt-RinketAl86PRB}, which causes a SF burst to form at the Fermi edge.  As a burst occurs, a significant population is depleted, resulting in a decreased Fermi energy.  Thus, as time goes on, the Fermi level moves toward the band edge continuously.  This results in a continuous line of SF emission at zero field and a series of sequential SF bursts in a magnetic field.

\newpage
\subsubsection{Theory of SF from Quantum Wells}

We use the semiconductor Bloch equations (SBEs) to study SF from a high-density $e$-$h$ plasma in the presence of many-body Coulomb interactions.  The usual form of the SBEs~\cite{HaugKoch04Book} is for a bulk semiconductor or a 2D electron gas, when the states can be labeled by a 3D or 2D wavevector $\vec{k}$. Here, we rederive SBEs following the same basic approximations but in a more general form, which accommodates the effects of a finite well width and the quantization of motion in a strong $B$.

We begin with a general Hamiltonian in the two-band approximation and $e$-$h$ representation,
\begin{equation}
\begin{split}
\hat{H} &=\sum_{\alpha}[(E_{g}^{0}+E_{\alpha}^{e})a_{\alpha}^{\dag}a_{\alpha}+E_{\alpha}^{h}b_{\bar{\alpha}}^{\dag}b_{\bar{\alpha}}] \\
  &+\frac{1}{2}\sum_{\alpha\beta\gamma\delta}(V_{\alpha\beta\gamma\delta}^{ee}a_{\alpha}^{\dag}a_{\beta}^{\dag}a_{\delta}a_{\gamma}+V_{\bar{\alpha}\bar{\beta}\bar{\gamma}\bar{\delta}}^{hh}b_{\bar{\alpha}}^{\dag} b_{\bar{\beta}}^{\dag}b_{\bar{\delta}}b_{\bar{\gamma}}\\&+2V_{\alpha\bar{\beta}\gamma\bar{\delta}}^{eh}a_{\alpha}^{\dag}b_{\bar{\beta}}^{\dag}b_{\bar{\delta}}a_{\gamma}               )
  -{\cal E}(t)\sum_{\alpha}(\mu_{\alpha}a_{\alpha}^{\dag}b_{\bar{\alpha}}^{\dag}+\mu_{\alpha}^{\ast}a_{\alpha}b_{\bar{\alpha}}),
\end{split}
\end{equation}
where $E_g^0$ is the unperturbed bandgap, $a_\alpha^\dagger$ and $b_{\bar{\alpha}}^\dagger$ are the creation operators for the electron state $\alpha$ and hole state $\bar{\alpha}$, respectively, ${\cal E}(t)$ is the optical field, $\mu_{\alpha}$ is the dipole matrix element, and $V_{\alpha\beta\gamma\delta}$ are Coulomb matrix elements, for example, $V^{ee}_{\alpha\beta\gamma\delta}$ = $\int d \vec{r}_1 \int d\vec{r}_2 \Psi^{e\ast}_\alpha(\vec{r}_1) \Psi^{e\ast}_\beta(\vec{r}_2) \frac{e^2}{\epsilon |\vec{r}_1 - \vec{r}_2|} \Psi^e_\gamma(\vec{r}_1) \Psi^e_\delta(\vec{r}_2)$. Here, we denote the hole state which can be recombined with a given electron state $\alpha$ optically by $\bar{\alpha}$, and assume that there is a one-to-one correspondence between them. For the interband Coulomb interaction, $V^{eh}_{\alpha\bar{\beta}\gamma\bar{\delta}} a_\alpha^\dagger b_{\bar{\beta}}^\dagger b_{\bar{\delta}} a_\gamma$ is the only nonzero matrix element due to the orthogonality between the Bloch functions of the conduction and valence bands~\cite{VaskoKuznetsov99Book}.  The electron and hole wave functions can be written as $\Psi^e_\alpha(\vec{r})$ = $\psi^e_\alpha(\vec{r}) u_{c0}(\vec{r})$ and $\Psi^h_{\bar{\alpha}}(\vec{r})$ = $\psi^h_{\bar{\alpha}}(\vec{r}) u^\ast_{v0}(\vec{r})$, respectively. In the problems we study here, the conduction band and valence band states connected by an optical transition always have the same envelope wave function, so we take $\psi^h_{\bar{\alpha}}(\vec{r})$ = $\psi^{e\ast}_\alpha(\vec{r})$. Then, the Coulomb matrix elements are related with each other through $V^{hh}_{\bar{\alpha}\bar{\beta}\bar{\gamma}\bar{\delta}} = V^{ee}_{\gamma\delta\alpha\beta}$ and $V^{eh}_{\alpha\bar{\beta}\gamma\bar{\delta}} = - V^{ee}_{\alpha\delta\gamma\beta}$, and we can drop the superscript by defining $V_{\alpha\beta\gamma\delta}$ $\equiv$ $V^{ee}_{\alpha\beta\gamma\delta}$.

Using the above Hamiltonian, we can obtain the equations of motion for the distribution functions $n_\alpha^e$ = $\langle a_\alpha^\dagger a_\alpha \rangle$ and $n_\alpha^h$ = $\langle b_{\bar{\alpha}}^\dagger b_{\bar{\alpha}} \rangle$, and the polarization $P_\alpha$ = $\langle b_{\bar{\alpha}} a_\alpha \rangle$. Using the Hartree-Fock approximation (HFA) and the random phase approximation (RPA), we arrive at the SBEs:
\begin{align}
\label{Eq::SBEs}
i\hbar {d\over d t} P_\alpha &= \left( E_g^0 + E^{eR}_\alpha + E^{hR}_\alpha \right) P_\alpha \nonumber \\
&+ \left( n^e_\alpha + n^h_\alpha - 1 \right) \left[ \mu_\alpha {\cal E}(t) + \sum_\beta V_{\alpha\beta\beta\alpha} P_\beta \right] \nonumber\\
& + \left. i\hbar {d\over d t} P_\alpha \right|_{\rm scatt} , \\
\hbar {d\over d t} n^e_\alpha &= - 2 ~\mathrm{Im} \left[ \left( \mu_\alpha {\cal E}(t) + \sum_\beta V_{\alpha\beta\beta\alpha} P_\beta \right) P_\alpha^\ast \right] \nonumber \\
&+ \left. \hbar {d\over d t} n^e_\alpha \right|_{\rm scatt} , \\
 \hbar {d\over d t} n^h_\alpha &= - 2 ~\mathrm{Im} \left[ \left( \mu_\alpha {\cal E}(t) + \sum_\beta V_{\alpha\beta\beta\alpha} P_\beta \right) P_\alpha^\ast \right] \nonumber\\
 &+ \left. \hbar {d\over d t} n^h_\alpha \right|_{\rm scatt} ,
 \end{align}
where $E^{eR}_\alpha$ = $\left( E^e_\alpha - \sum_\beta V_{\alpha\beta\beta\alpha} n^e_\beta \right)$ and $E^{hR}_\alpha$ = $\left( E^h_\alpha - \sum_\beta V_{\alpha\beta\beta\alpha} n^h_\beta \right)$ are the renormalized energies, and the scattering terms account for higher-order contributions beyond the HFA and other scattering processes such as longitudinal-optical phonon scattering.

\begin{figure}[hbtp]
\centering
\includegraphics[scale = 0.59]{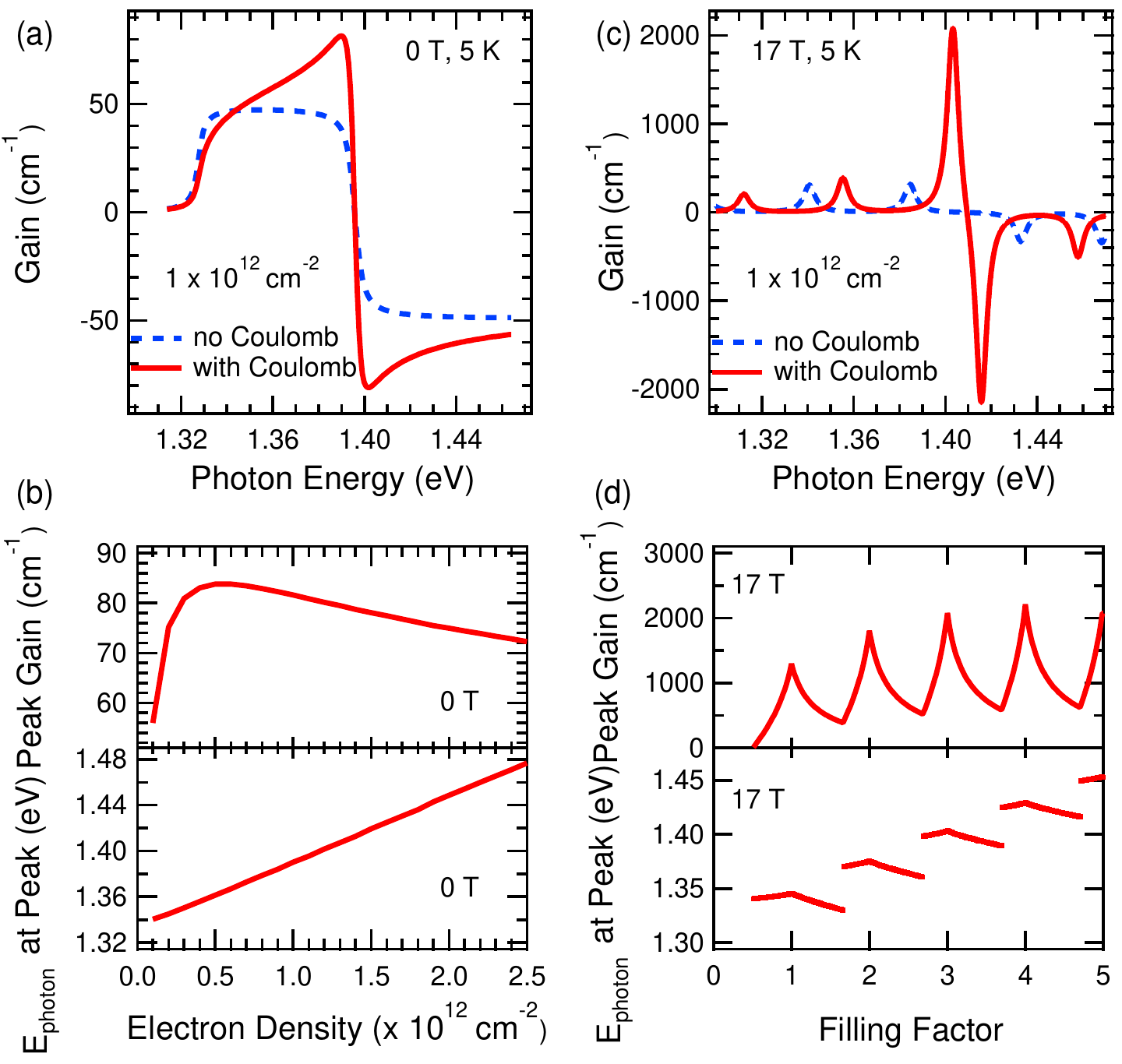}
\caption{Theoretical calculations of Coulomb-induced many-body enhancement of gain at the Fermi energy at zero magnetic field and 17\,T.
(a)~Gain spectrum for the InGaAs sample without a magnetic field with (solid line) and without (dashed line) considering Coulomb effects; (b)~Peak gain (upper panel) and peak gain energy (lower panel) as a function of $e$-$h$ density at zero magnetic field. (c)~Calculated gain spectrum in a magnetic field of 17\,T with (solid line) and without (dash line) considering Coulomb effects; (d)~Peak gain (upper panel) and peak gain energy (lower panel) at 17\,T as a function of filling factor. Reproduced (adapted) with permission from~\cite{KimetAl13SR}. Copyright 2013, Nature Publishing Group.}
\label{GainTheory}
\end{figure}

These equations, together with Maxwell's equations for the electromagnetic field, can be applied to study the full nonlinear dynamics of interaction between the $e$-$h$ plasma and radiation. Here, we derive the gain for given carrier distributions $n_\alpha^e$ and $n_\alpha^h$, which was used to plot Fig.\,\ref{GainTheory}. Assuming a monochromatic and sinusoidal time dependence for the field ${\cal E}(t)$ = ${\cal E}_0 e^{-i\omega t}$ and the polarization $P_\alpha(t)$ = $P_{0\alpha} e^{-i\omega t}$, we can find $P_\alpha$ from Eq.~(\ref{Eq::SBEs}) and define the quantity $\chi_\alpha(\omega)$ = $P_{0\alpha}/{\cal E}_0$, which satisfies
\begin{eqnarray}
\label{Eq::chi}
\chi_\alpha(\omega) = \chi^0_\alpha(\omega) \left[ 1 + {1 \over \mu_\alpha} \sum_{\beta} V_{\alpha\beta\beta\alpha} \chi_\beta(\omega) \right] ,
\end{eqnarray}
where
\begin{eqnarray}
\label{Eq::chi0}
\chi^0_\alpha(\omega) = \frac{\mu_\alpha \left( n^e_\alpha + n^h_\alpha - 1 \right)} {\hbar\omega - \left( E_g^0 + E^{eR}_\alpha + E^{hR}_\alpha \right) + i\hbar\gamma_\alpha} .
\end{eqnarray}
Here, we have written the dephasing term phenomenologically as $d P_\alpha / d t |_{\rm scatt} = - \gamma_\alpha P_\alpha$. The optical susceptibility is then
\begin{eqnarray}
\label{Eq::chiw}
\chi(\omega) = {1 \over V} \sum_\alpha \mu_\alpha^\ast \chi_\alpha(\omega) ,
\end{eqnarray}
where $V$ is the normalization volume. The gain spectrum is given by~\cite{HaugKoch04Book}
\begin{eqnarray}
\label{Eq::gainw}
g(\omega) = \frac{4\pi\omega}{n_b c} \mathrm{Im} [\chi(\omega)] ,
\end{eqnarray}
where $n_b$ is the background refractive index, and $c$ is the speed of light.  We use the above general results to analyze optical properties under different conditions.

In a QW of thickness  $L_{\rm w}$, the envelope functions for  electrons and holes are  $\psi^{e,h}_{n,\vec{k}}(\vec{r})$ = $\varphi_n(z)\exp \left(i \vec{k} \cdot \vec{\rho} \right)/\sqrt{A}$, where $\vec{\rho}$ = $(x,y)$, $\varphi_n(z)$ is the envelope wave function in the growth direction for the $n$-th subband, and $A$ is the normalization area.  To calculate the Coulomb matrix element $V_{\alpha\beta\beta\alpha}$, we define $\tilde{V}_{\alpha\beta}$ $\equiv$ $V_{\alpha\beta\beta\alpha}$ and put $\alpha$ = $\left\{ n, \vec{k}, s \right\}$, $\beta$ = $\left\{ n', \vec{k}', s' \right\}$, where $s$ denotes the spin quantum index. Then one gets
\begin{eqnarray}
\tilde{V}_{n,\vec{k},s;n',\vec{k}',s'} = V^{2D}(q) F_{n n' n' n}(q) \delta_{s s'} ,
\end{eqnarray}
where $q$ = $|\vec{q}|$ = $|\vec{k}-\vec{k}'|$, $V^{2D}(q)$ = $2\pi e^2 / \epsilon A q$, $\epsilon$ is the dielectric function, and the form factor $F_{n n' n' n}(q)$ is defined as
\begin{align}
\label{Eq::Formfactor}
&\phantom{{}={}} F_{n1, n2, n3, n4}(q) \nonumber \\
&= \int d z_1 \int d z_2 \varphi_{n1}^\ast(z_1) \varphi_{n2}^\ast(z_2) \exp \left( - q \left| z_1 - z_2 \right| \right) \nonumber \\
&\times  \varphi_{n3}(z_1) \varphi_{n4}(z_2) .
\end{align}
Throughout this section, we assume that only the lowest conduction and valence subbands are occupied. In this case, we can define $\tilde{V}(q)$ = $V^{2D}(q) F_{1111}(q)$. The dielectric function $\epsilon(\vec{q},\omega)$, which describes the screening of the Coulomb potential, is given by the Lindhard formula for a pure 2D case \cite{HaugKoch04Book}; it can be generalized to the quasi-2D case as
\begin{eqnarray}
\label{Eq::Lindhard}
\epsilon(\vec{q},\omega) = 1 + \tilde{V}(q) \left( \Pi_e(\vec{q},\omega) + \Pi_h(\vec{q},\omega) \right) ,
\end{eqnarray}
where $\Pi_{e(h)}(\vec{q},\omega)$ is the polarization function of an electron or hole, which is given by
\begin{eqnarray}
\Pi(\vec{q},\omega) = 2 \sum_{\vec{k}} \frac{n_{\vec{k}+\vec{q}} - n_{\vec{k}}}{\omega + i 0^+ - E_{\vec{k}+\vec{q}} + E_{\vec{k}}} .
\end{eqnarray}
Here, we dropped the subscripts $e$ or $h$, $n_{\vec{k}}$ is  the distribution function, the factor of 2 accounts for the summation over spin, and the spin index is suppressed. For simplicity, we will choose the static limit, namely, $\omega$ = 0.

Given the dielectric function $\epsilon(q,0)$, the screened Coulomb matrix element is $\tilde{V}_s(q)$ = $\tilde{V}(q)/\epsilon(q,0)$. For simplicity, we will still write it as $\tilde{V}(q)$. Applying Eq.\ (\ref{Eq::chi}) to the case above, we get the equation for $\chi_{\vec{k}}(\omega)$:
\begin{eqnarray}
\label{Eq::chiQW}
\chi_{\vec{k}}(\omega) = \chi^0_{\vec{k}}(\omega) \left[ 1 + {1 \over \mu_{\vec{k}}} \sum_{\vec{k}'} \tilde{V}\left( \left| \vec{k}-\vec{k}' \right| \right) \chi_{\vec{k}'}(\omega) \right] ,
\end{eqnarray}
where $\chi^0_{\vec{k}}(\omega)$ becomes
\begin{eqnarray}
\chi^0_{\vec{k}}(\omega) = \frac{\mu_{\vec{k}} \left( n^e_{\vec{k}} + n^h_{\vec{k}} - 1 \right)} {\hbar\omega - \left( E_g^0 + E^{eR}_{\vec{k}} + E^{hR}_{\vec{k}} \right) + i\hbar\gamma_{\vec{k}}} .
\end{eqnarray}

To solve Eq. (\ref{Eq::chiQW}), we notice that $\chi^0_{\vec{k}}(\omega)$ does not depend on the direction of $\vec{k}$, so $\chi_{\vec{k}}(\omega)$ will not depend on it, either. Then, after converting the summation in Eq.~(\ref{Eq::chiQW}) into the integral, the integration over the azimuthal angle is acting on $\tilde{V} ( | \vec{k}-\vec{k}' | )$ only. If we define
\begin{eqnarray}
 \tilde{V}\left(k,k' \right) = {1\over 2\pi} \int_0^{2\pi} d \phi \tilde{V}\left( \sqrt{k^2+k^{'2} - 2 k k' \cos \phi} \right) ,
\end{eqnarray}
then Eq.~(\ref{Eq::chiQW}) can be written as
\begin{eqnarray}
\chi_k(\omega) = \chi^0_k(\omega) \left[ 1 + {A \over 2 \pi \mu_k} \int_0^\infty k' d k'  \tilde{V}\left( k, k' \right) \chi_k'(\omega) \right] .
\end{eqnarray}
After discretizing the integral, we have a system of linear equations for $\chi_k(\omega)$, which can be solved by using LAPACK~\cite{LAPACK-guide}. The band structure for our sample consisting of undoped 8-nm In$_{0.2}$Ga$_{0.8}$As wells and 15-nm GaAs barriers on a GaAs substrate is calculated using the parameters given by Vurgaftman, Meyer, and Ram-Mohan~\cite{VurgaftmanetAl01JAP}. The strain effect is included using the results of Sugawara {\it et al}.~\cite{SugawaraetAl93PRB}. Examples of calculated gain spectra are shown in Figs.\,\ref{GainTheory}(a) and \ref{GainTheory}(b).

For a QW structure in a strong perpendicular $B$, the electronic states are fully quantized. Considering only the lowest subband in the QW, the equation for the susceptibility is written as
\begin{eqnarray}
\chi_{\nu,s} = \chi^0_{\nu,s} \left[ 1 + {1 \over \mu_{\nu,s}} \sum_{\nu'} V_{\nu,\nu'} \chi_{\nu',s}\right] ,
\end{eqnarray}
where $\nu$ is the Landau level index, $s$ is the spin index, and $V_{\nu,\nu'}$ is the Coulomb matrix element given by
\begin{align}
V_{\nu,\nu'} &= \frac{e^2}{2\pi\epsilon} \int_0^{2\pi} d\theta \int_0^{\infty} d q \nonumber \\
& \times \left| \int d x e^{i q x \cos\theta} \phi_{\nu} (x) \phi_{\nu'}^\ast(x + q a_H^2 \sin\theta) \right|^2 ,
\end{align}
where $\phi_{\nu}(x)$ is the $x$-dependent part of the wave function of the $\nu$-th Landau level and $a_H^2$ = $\hbar c/e B$. The renormalized electronic energies in the expression for $\chi^0_{\nu,s}$ are
\begin{eqnarray}
E_{\nu,s}^{eR} = E_{\nu,s}^e - \sum_{\nu'} V_{\nu,\nu'} n_{\nu'}^e  ,
\end{eqnarray}
and a similar equation holds for holes. The gain is calculated as
\begin{eqnarray}
\label{gainmag}
g(\omega) = \frac{4\pi\omega}{n_b c} \frac{1}{\pi a_H^2} \mathrm{Im} \left[ \sum_{\nu} \mu_{\nu,s}^\ast \chi_{\nu,s} \right] .
\end{eqnarray}
An example of the calculated gain for $B$ = 17\,T is shown in Figs.\,\ref{GainTheory}(c) and \ref{GainTheory}(d).

\subsection{Superradiant Decay of Coherent Cyclotron Resonance in Ultrahigh-Mobility Two-Dimensional Electron Gases}
\label{SR-CR}

Solid-state Dicke SR arising from extended states can also happen in {\em intraband} transitions in semiconductors, such as cyclotron resonance (CR)~\cite{ZhangetAl14PRL} and intersubband transitions~\cite{LaurentetAl15PRL} in QWs. In this section, we deal with SR of CR.  Specifically, we show that superradiant decay can dominate the nonequilibrium dynamics of interacting electrons in a Landau-quantized, high-mobility two-dimensional electron gas (2DEG). The coherence in such a system is created through resonant excitation by an external light field, as in the case of excitonic SR (Section \ref{xSR}), as opposed to the spontaneously emerged coherence in the case of SF.

\begin{figure}[t]
\centering
\includegraphics[scale=0.7]{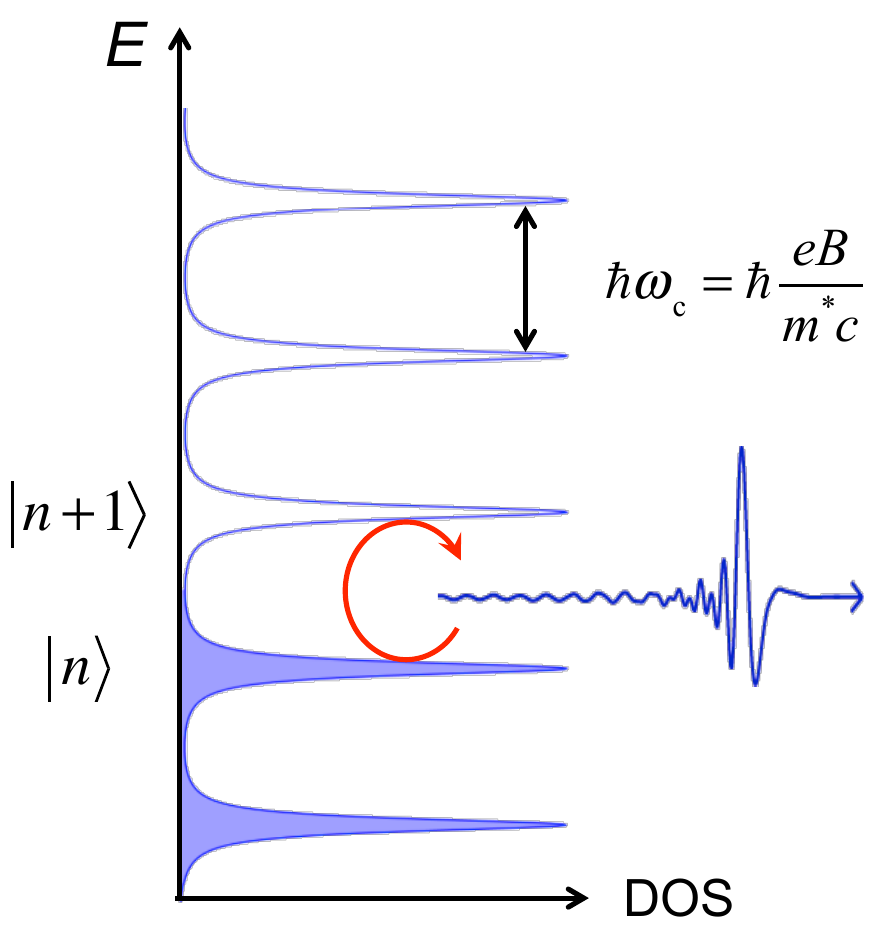}
\caption{Coherent THz pulse creates a superposition of adjacent Landau levels with massive degeneracy. The Landau level spacing equals $\hbar\omega_\mathrm{c}$, where $\omega_\mathrm{c}$ is the cyclotron frequency. The free induction decay of such a superposition state can be observed after the excitation pulse.}
\label{CR_LL}
\end{figure}

From a quantum mechanical point of view, CR is the evolution of a coherent superposition of adjacent Landau levels (LLs), with massive degeneracy, as schematically shown in Fig.~\ref{CR_LL}.  How rapidly the coherence of this many-body superposition state decays has not been well understood.  Even though the CR frequency, $\omega_\textrm{c}$, is immune to many-body interactions due to Kohn's theorem~\cite{Kohn61PR}, the decoherence of CR can be affected by electron-electron interactions. Theoretical studies predicted that the linewidth of CR should oscillate with the LL filling factor since the screening capability (i.e., the density of states at the Fermi energy) of a 2DEG oscillates with the filling factor~\cite{Ando75JPSJ,Ando77JPSJ,DasSarma81PRB,LassnigGornik83SSC,FoglerShklovskii98PRL}. However, despite several decades of experimental studies of CR in 2DEGs using continuous-wave spectrometers~\cite{EnglertetAl83SSC,SchlesingeretAl84PRB,HeitmannetAl86PRB,EnsslinetAl87PRB,SeidenbuschGornik87PRB,BatkeetAl88PRB,KonoetAl94PRB}, no clear evidence for the predicted CR linewidth oscillations has been obtained for high-mobility, high-density samples.

Here, we present a systematic study on CR decoherence in high-mobility 2DEGs by using time-domain THz magneto-spectroscopy. We found that the polarization decay rate at the CR, $\Gamma_\mathrm{CR}$ ($\equiv$ $\tau_\mathrm{CR}^{-1}$) increases linearly with the electron density, $n_\text{e}$, which is the signature of SR (or radiation damping)~\cite{Dicke54PR,Haken84Book,BloembergenPound54PR}. Namely, the decay of CR is dominated by a cooperative radiative decay process, which is much faster than any other phase-breaking scattering processes for an individual electron. This model explains the absence of CR linewidth oscillations with respect to the filling factor and a low temperature saturation of the CR decay time, $\tau_\text{CR}$.

Two samples of modulation-doped GaAs QWs were used. Sample 1 had an electron density $n_\text{e}$ and mobility $\mu_\text{e}$ of 1.9 $\times$ 10$^{11}$\,cm$^{-2}$ and 2.2 $\times$ 10$^6$\,cm$^2$/Vs, respectively, in the dark, while after illumination at 4\,K they changed to 3.1 $\times$ 10$^{11}$\,cm$^{-2}$ and 3.9 $\times$ 10$^6$\,cm$^2$/Vs; intermediate $n_\text{e}$ values were achieved by careful control of illumination times.  Sample 2 had $n_\text{e} =$ 5 $\times$ 10$^{10}$\,cm$^{-2}$ and $\mu_\text{e} =$ 4.4 $\times$ 10$^6$\,cm$^2$/Vs.

\subsubsection{Observation of Superradiant Decay of CR}

\begin{figure}
\includegraphics[scale=0.4]{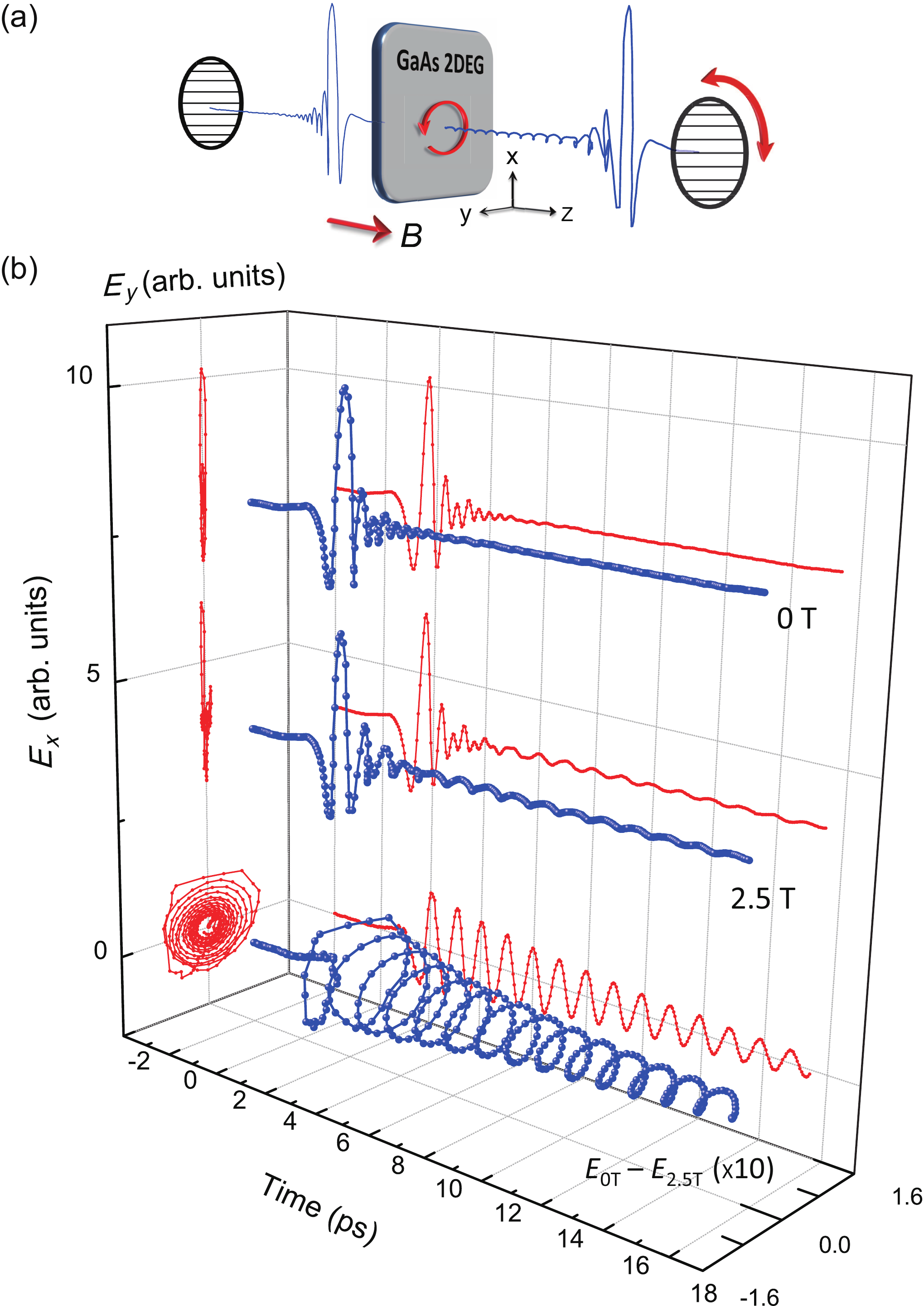}
\caption{(a)~A schematic of the polarization-resolved THz magneto-transmission experiment in the Faraday geometry.  (b)~Coherent cyclotron resonance oscillations in the time domain.  Each blue dot represents the tip of the THz electric field at a given time.  The red traces are the projections of the waveforms onto the $E_x$-$t$ and $E_x$-$E_y$ planes.  The bottom trace is the difference between the top (0\,T) and middle (2.5\,T) traces. Reproduced (adapted) with permission from~\cite{ZhangetAl14PRL}. Copyright 2014, American Physical Society.}
\label{CR_time}
\end{figure}

We performed time-domain THz magneto-spectroscopy~\cite{WangetAl07OL,ArikawaetAl11PRB} experiments. The incident THz beam was linearly polarized by the first polarizer, and by rotating the second polarizer, the transmitted THz field was measured in both $x$- and $y$-directions [Fig.\,\ref{CR_time}(a)]. Figure~\ref{CR_time}(b) shows transmitted THz waveforms in the time domain.  Each blue dot represents the tip of the THz electric field, $\vec{E}$ = ($E_x$,$E_y$), at a given time.  The red traces are the projections of the waveforms onto the $E_x$-$t$ plane and $E_x$-$E_y$ plane.  The top and middle traces show the transmitted THz waveforms at 0\,T and 2.5\,T, respectively. The 2.5\,T trace contains long-lived oscillations with circular polarization.  The bottom trace is the difference between the two, $E_\mathrm{0\,T}(t)-E_\mathrm{2.5\,T}(t)$, which is the free induction decay signal of CR. Hence, CR decay time, $\tau_\mathrm{CR}$, can be accurately determined through time-domain fitting with $A\exp(-t/\tau_\mathrm{CR})\cdot\sin(\omega_\mathrm{c}t+\phi_0)$, where $A$ and $\phi_0$ are the CR amplitude and the initial phase, respectively.

\begin{figure}
\includegraphics[width=\linewidth]{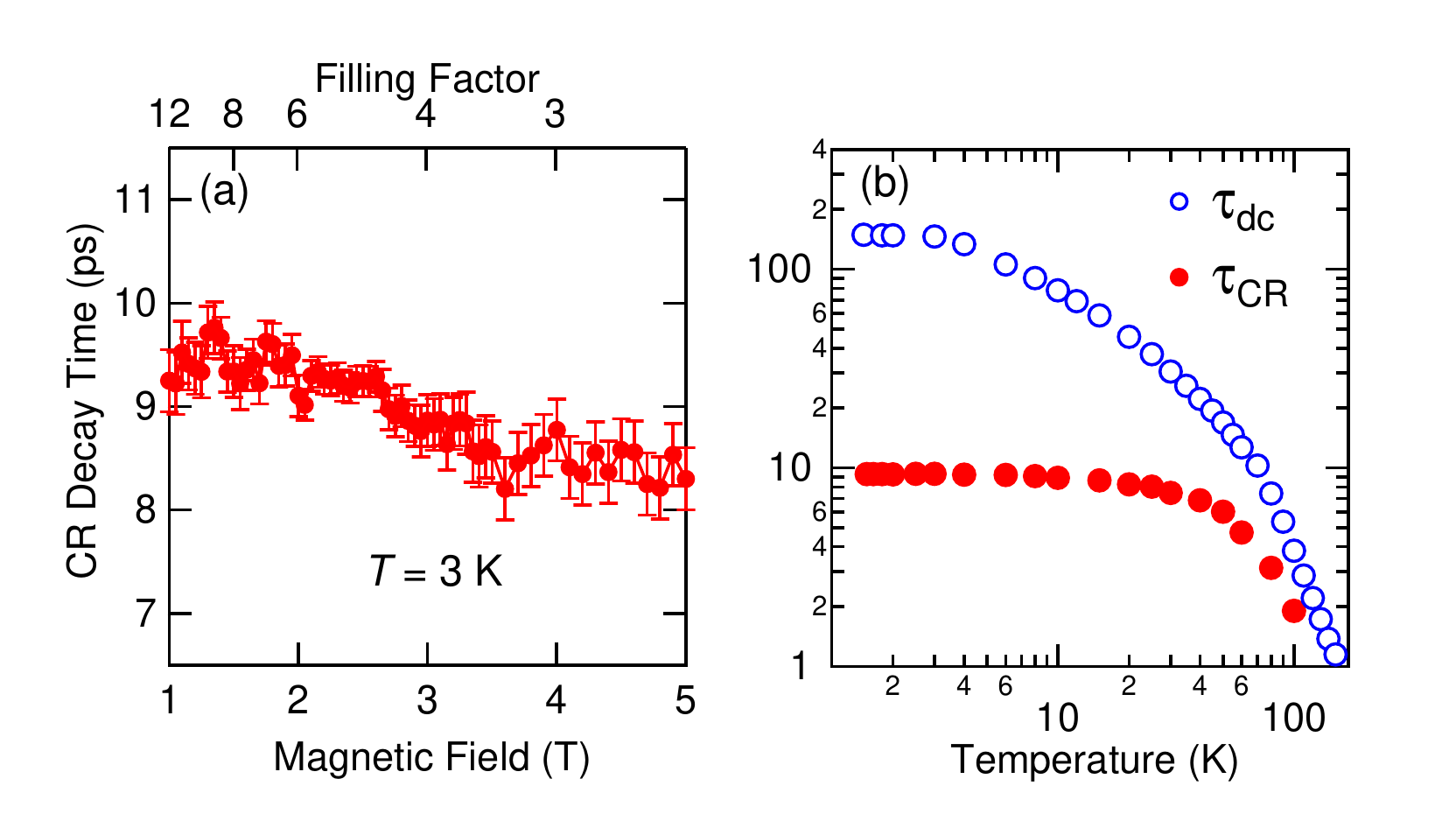}
\caption{(a)~Magnetic field dependence of $\tau_\mathrm{CR}$ at 3\,K. (b)~Temperature dependence of $\tau_\mathrm{CR}$ at 2.5\,T.  All the data are for Sample~1. Reproduced (adapted) with permission from~\cite{ZhangetAl14PRL}. Copyright 2014, American Physical Society.}
\label{CR_BT}
\end{figure}

\begin{figure}
\includegraphics[width=1.05\linewidth]{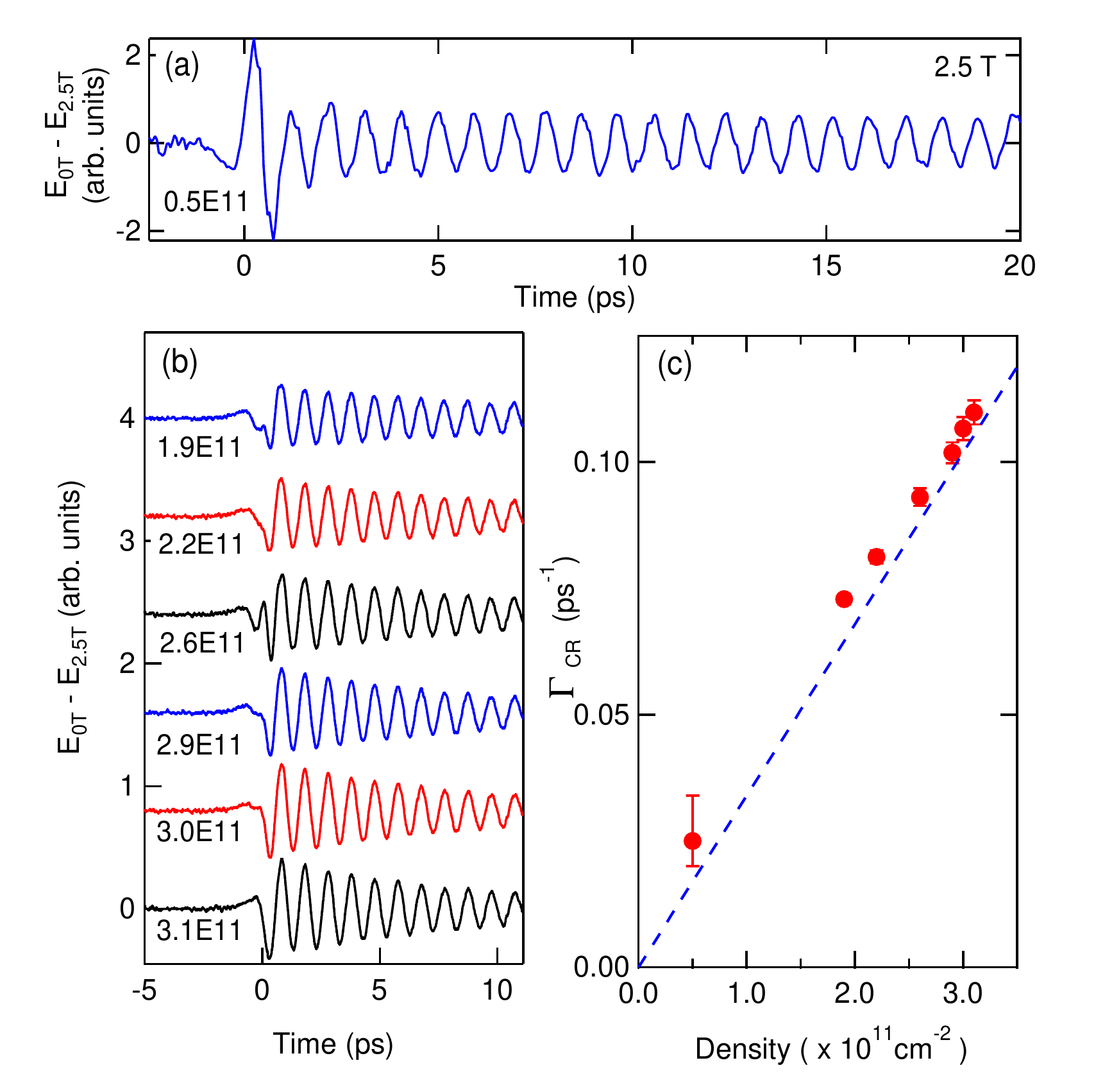}
\caption{(a)~Low-density sample (Sample 2) exhibiting the longest $\tau_\mathrm{CR}$ of $\sim$40\,ps. (b)~CR oscillations in Sample 1 with different densities by controlling the illumination time. (c)~Decay rate as a function of density. Blue solid circle: Sample 2.  Red solid circles: Sample 1.  The blue dashed line represents Eq.~(\ref{Gamma-SR}) with no adjustable parameter. Reproduced (adapted) with permission from~\cite{ZhangetAl14PRL}. Copyright 2014, American Physical Society.}
\label{CR_Density}
\end{figure}

Figure~\ref{CR_BT}(a) shows the magnetic field dependence of $\tau_\mathrm{CR}$; $\tau_\mathrm{CR}$ slightly decreases with increasing $B$.  Figure~\ref{CR_BT}(b) shows that $\tau_\mathrm{CR}$ saturates at $\sim$9.5\,ps when $T \lesssim$ 10\,K. The values were much shorter than the DC scattering time, $\tau_{\rm DC}$ = $m^*\mu_\text{e}/e$, of the same samples at the same temperature.  Furthermore, no correlation was found between $\tau_\mathrm{CR}$ and $\tau_{\rm DC}$; on the other hand, $\tau_\mathrm{CR}$ showed strong correlation with $n_\text{e}$.  As $n_\text{e}$ was increased, $\tau_\mathrm{CR}$ was found to decrease in a clear and reproducible manner. As shown in Figs.\,\ref{CR_Density}(a) and \ref{CR_Density}(b). The low-density sample (Sample 2) exhibited the longest $\tau_{\rm CR}$ value of $\sim$40\,ps.  Figure~\ref{CR_Density}(c) shows that the decay rate, $\Gamma_\mathrm{CR}$, increases linearly with $n_\text{e}$, which, as described below, is consistent with superradiant decay of CR.

A qualitative picture of superradiant decay of CR is as follows. A coherent incident THz pulse induces a polarization in the 2DEG, i.e., macroscopic coherence as a result of individual cyclotron dipoles oscillating in phase.  The resulting free induction decay of polarization occurs in a superradiant manner, much faster than the dephasing of single oscillators. The SR decay rate, $\Gamma_\mathrm{SR}$, is roughly $N$ times higher than the individual radiative decay rate, where $N \sim n_\text{e} \lambda^2 \sim n_\text{e}/{\omega_\mathrm{c}}^2$ is the number of electrons within the transverse coherence area of a radiation wavelength $\lambda$. The spontaneous emission rate for individual CR, which is a quantum harmonic oscillator, is proportional to ${\omega_\mathrm{c}}^2$. Therefore, the collective radiative decay rate $\Gamma_\mathrm{SR}$ has no explicit $\omega_\mathrm{c}$ or $B$ dependence. In an ultraclean 2DEG, $\Gamma_\mathrm{SR}$ can be higher than the rates of all other phase-breaking scattering mechanisms.  This scenario explains not only the $n_e$ dependence of $\tau_\mathrm{CR}$ but also its weak $B$ dependence [Fig.\,\ref{CR_BT}(a)] as well as the saturation of $\tau_\mathrm{CR}$ at low temperature.

\subsubsection{Theory of Superradiant Decay of Coherent CR}

We developed a quantum mechanical model for THz excitation and coherent CR emission of a 2DEG in a perpendicular $B$~\cite{ZhangetAl14PRL}, based on the master equation for the density operator in the coordinate representation, $d\hat{\rho}/dt = - (i/\hbar) [ \hat{H},\hat{\rho} ] + \hat{R}(\hat{\rho})$,
%
%
where $\hat{R}(\hat{\rho})$ is the relaxation operator. Here, the Hamiltonian for an electron of mass $m^*$ in a confining potential $U(\bf{r})$ interacting with an optical and magnetic field described by the vector potential $\vec{A} = \vec{A}_\mathrm{opt} + \vec{A}_{B}$ is
 \begin{equation}
 \label{ham}
\hat{H} = \frac{\hat{p}^2}{2m^*} +U({\bf r}) - \frac{e}{2m^*c} \left(\vec{A}\cdot \hat{\vec{p}} + \hat{\vec{p}} \cdot \vec{A} \right) + \frac{e^2}{2m^*c^2} |\vec{A}|^2,
\end{equation}
where $\hat{\vec{p}} = - i \hbar \nabla$.

As shown in \cite{ZhangetAl14PRL},  the density matrix equations including both the electric field of the excitation pulse,  $\vec{E}_0 = (E_{0x}(t), 0, 0)$ and the field radiated by the circularly polarized electron current $j_+ = j_x - ij_y$, result in the following equation of motion for $j_+$:
\begin{equation}
\label{cur2}
{dj_+ \over dt} + (i \omega_c + \Gamma_\mathrm{CR}) j_+  = \alpha E_{0x}(t),
\end{equation}
where $\alpha$ = $\omega_p^2/4\pi$ and $\omega_p$ = $(4 \pi e^2 \overline{\hat{\rho}}/m^*)^{1/2}$ is the plasma frequency.  The CR decay rate, $\Gamma_\mathrm{CR}$, includes the collective radiative contribution proportional to $n_\text{e}$
\begin{equation}
\label{Gamma}
 \Gamma_\mathrm{CR} = 2\gamma_\mathrm{\perp} + \Gamma_\mathrm{SR},
\end{equation}
where
\begin{equation}
\label{Gamma-SR}
\Gamma_\mathrm{SR} = \frac{4\pi e^2n_e}{m^*(1+n_\mathrm{GaAs})c}.
\end{equation}
Here, $ \gamma_\mathrm{\perp}$ is the relaxation rate of the off-diagonal component of the density matrix and $n_\mathrm{GaAs}$ = 3.6 is the refractive index of the GaAs substrate. As shown by the dashed line in Fig.\,\ref{CR_Density}(c), Eq.\,(\ref{Gamma-SR}) reproduces the observed linear $n_\text{e}$ dependence of $\Gamma_\mathrm{CR}$ \emph{without any adjustable parameters}, strongly supporting the notion that superradiant decay (radiation damping) dominates the CR decay process in these high-$\mu_\text{e}$ samples.
Equation~(\ref{Gamma}) also allows us to determine $2 \gamma_\mathrm{\perp}$~as $\Gamma_\mathrm{CR} - \Gamma_\mathrm{SR}$.  In particular, we interpret the small but non-negligible $B$-dependence of $\Gamma_\mathrm{CR}$ shown in Fig.\,\ref{CR_Density}(c) to be the $B$-dependence of $\gamma_\mathrm{\perp}$.

Furthermore, $\Gamma_\mathrm{SR}$ is inversely proportional to the carrier effective mass, $m^*$, indicating that SR decay of CR is stronger in narrow bandgap semiconductors, e.g., InSb. In a 2D hole gas (2DHG), on the other hand, weaker SR decay is expected due to the much heavier effective mass; this was confirmed by recent experiments in a high-mobility GaAs 2DHG, where $\Gamma_\mathrm{SR}$ was comparable with the intrinsic dephasing rate $\gamma_\mathrm{\perp}$~\cite{KamarajuatAl15APL}. In general, SR decay of CR could also happen in 3D semiconductors, but the scattering rate there is expected to be higher because of a continuous spectrum of carriers.  

To reveal the intrinsic phase-breaking scattering processes of CR in high-mobility, high-density 2DEGs, suppression of SR decay is required. The spontaneous  decay rate  can be modified by changing the dielectric environment or putting the sample into a high-$Q$ cavity. Especially in the strong light-matter coupling regime, the reversible emission/absorption leads to the exchange of energy between light and matter, and thus, the radiation decay is suppressed. Experimentally, such a situation has been achieved by strongly coupling CR to plasmons~\cite{AndreevetAl14APL} or cavity photons~\cite{ZhangetAl16arXiv}. 






\newpage
\section{Summary}

We reviewed the current state of the field of cooperative spontaneous emission, first put forward by Dicke~\cite{Dicke54PR}, in the novel context of nonequlilibrium condensed matter systems.  Unlike the corresponding concepts in traditional atomic and molecular gases, these phenomena acquire different appearances in solid-state environments because of the inherently fast dephasing and strong Coulomb interactions.  Excitonic interactions and coupling between electrons and holes are particularly important both in superradiant decays and superfluorescent bursts~\cite{NoeetAl12NP}.  Massively Fermi-degenerate electrons and holes, which would never occur in atomic-like systems, can lead to many-body enhancement of gain, which induces preferential production of a superfluorescent burst at the Fermi edge~\cite{KimetAl13SR}.  This is still a rapidly progressing field of research, expanding to encompass more and more nontraditional physical situations for SR and SF, such as plasmon excitations~\cite{SonnefraudetAl10ACSNano,Martin-CanoetAl10NL} and exciton-plasmon coupling~\cite{ChenetAl08OL,TeperikDegiron12PRL}, with unique solid-state cavities to create nonintuitive many-body playgrounds~\cite{ZhangetAl16arXiv,SuetAl13PRL,LeymannetAl15PRA}.

\section*{Acknowledgments}

We acknowledge support from the National Science Foundation through Grant Nos.\ DMR-1006663, DMR-1310138, and ECS-0547019. 
Y.W.\ and A.B.\ were supported in part by  the Air Force Office for Scientific Research through grant FA9550-15-1-0153.  We thank S.\ A.\ McGill for assistance with measurements performed at the National High Magnetic Field Laboratory, G.\ S.\ Solomon for providing us with the InGaAs/GaAs quantum well sample used in the SF study (Section \ref{SF-QW}), and J.\ L.\ Reno, W.\ Pan, J.\ D.\ Watson, and M.\ J.\ Manfra for the growth of the ultrahigh-mobility 2DEG samples for our CR studies (Section \ref{SR-CR}).






\end{document}